\documentclass[12pt]{article}
\usepackage{amsmath}
\usepackage{amsfonts}
\usepackage{slashed}
\usepackage{arydshln}

\usepackage[body={16cm,22.5cm}]{geometry}

\newcommand{\be}{\begin{equation}}
\newcommand{\ee}{\end{equation}}
\newcommand{\bea}{\begin{eqnarray}}
\newcommand{\eea}{\end{eqnarray}}

\newcommand{\eq}[1]{(\ref{#1})}
\newcommand{\Eq}[1]{Eq.~(\ref{#1})}
\newcommand{\nn}{\nonumber}

\def\H{\Phi}

\def\lra#1{\overset{\text{\scriptsize$\leftrightarrow$}}{#1}}

\def\stw{s_{w}}
\def\ctw{c_{w}}
\def\ttw{t_{w}}

\def\cA{ {\cal A } }
\def\cZ{ {\cal Z } }
\def\cW{ {\cal W } }
\def\cG{ {\cal G } }
\def\cV{ {\cal V } }

\begin{document}

\title{\bf  An Effective Guide to \\
Beyond the Standard Model Physics}
\author{ \bf{Eduard Mass{\'o}}\\
\small{ \it Grup de F{\'\i}sica Te{\`o}rica and Institut de F{\'\i}sica d'Altes Energies,}\\
\small{  \it Universitat Aut{\`o}noma de Barcelona, 08193 Bellaterra, Spain}}

\date{}

\maketitle

\vspace*{1cm}
\centerline{\large \bf Abstract}
\vspace*{.5cm}

Effective Lagrangians with dimension-six operators are widely used to analyse
Higgs and other electroweak data. We show how to build a basis of operators
such that each operator is linked to a coupling which is well measured or will be in the future. 
We choose a set of couplings such that the correspondence is one-to-one.
The usual arbitrary coefficients in front of operators can be constrained quite
directly using experimental data.
In our framework, some important features of the Lagrangian are  transparent. For example, 
one can clearly see the presence or absence of correlations among measurable
quantities. This may be a useful guide when searching for physics beyond the
Standard Model.

\vspace*{1cm}

\section{Introduction}

The discovery of the Higgs boson \cite{Chatrchyan:2012twa,ATLAS:2012ad} at the LHC
has given a significant boost to the  Standard Model (SM) of particle physics: the Higgs $h$ completes
the spectrum of the model and  its properties are fairly consistent
with  the SM predictions. However, we still think that the SM is not the final
theory of particle physics and a pressing issue is whether there will be some sign of beyond the SM (BSM)
physics once the Higgs properties are known with more precision.
The analysis can be done different high-energy models, but a general tool that is now extensively used is the effective Lagrangian approach
\cite{Willenbrock:2014bja}-\cite{Contino}.

The effective Lagrangian framework allows a model independent scrutiny of electroweak data.
In this paper we will adopt its linear version, suitable when the scale of BSM
physics $\sim \Lambda$ is much higher than the electroweak scale (as indicated by LHC results), and the BSM degrees of freedom have been integrated out, leaving below 
$\Lambda$ the SM degrees of freedom, including   $h$, which we assume is part of an electroweak doublet.
The resulting effective theory at energies below $\Lambda$ can be described by a Lagrangian formed
by gauge-invariant operators formed by the Higgs, gauge bosons and fermion fields, in an expansion according to the dimension $d$ of the operators. The SM Lagrangian 
contains all $d=4$ terms.
We assume that we can truncate the expansion of the effective Lagrangian
at the dominant $d=6$ operators,\footnote{The scale of $d=5$ operators related to $L$ non-conservation is at much higher scales.}
\begin{equation}
\label{Leff1}
{\cal L}_{6} = \sum_i \frac{c_i}{\Lambda^2} \, {\cal O}_i  \ .
\end{equation}
The sum over $i$ runs over a basis $\{ {\cal O}_i \}$ in the $d=6$ gauge-invariant operator space.
One of the guiding principles to elect a particular basis should be that it simplifies the path to get to
the desired goals.

To motivate our work we start with the well-know fact that the structure of the SM Lagrangian implies
relations among couplings. Consider for instance the  SM Lagrangian term
\be
 m^2_W  \left( W^{+ \mu} W^-_\mu + \frac{1}{2 \, \ctw^2}  Z^{\mu} Z_\mu \right)  \, 
 \left(1 +  \frac{2h}{v}  + \frac{h^2}{v^2} \right)  \ ,
    \label{smRelation1}
\ee
where $h$ stands for the physical Higgs boson, and $v$, $\ctw=\cos \theta_w$, etc., are defined in Appendix A.
From \eq{smRelation1}, we see for example that the couplings $hWW$ and $hZZ$ are related,
 through the expected custodial-preserving form.
The couplings $h^2WW$-$h^2ZZ$, $hWW$-$h^2WW$,
and $hZZ$-$h^2ZZ$ are also related.
 These relations hold exactly at tree level, but they are slightly changed by radiative corrections. 
In this paper we are interested in the changes induced by $d=6$ operators.
It is well-known that the modifications coming from ${\cal L}_{6}$
are not arbitrary. Indeed there are correlations amongst the induced modifications
which stem from the fact that gauge-invariance restricts the form of the operators,
and that not all the operators one can write are independent. However, as we now explain, to see clearly
those correlations in complete generality may be difficult.

One uses (\ref{Leff1})
to predict modifications to couplings, which we denote generically by $a$. The experimental values
for the different $a$'s, obtained from measurements, constrain the $c_i$'s.
However, in general, a $d=6$ operator in (\ref{Leff1}) 
contributes to several couplings $a$, and in turn each coupling $a$ gets contributions
 from a certain number of operators. Thus, the connection among $c_i$'s and $a$'s may be involved.
 When doing a numerical fit of all $c_i$  using observational data,
 one may loose sight of the correlations above mentioned
and of some other general properties of the physics of effective Lagrangians.
 It would be desirable to have a basis where the relations among $c_i$'s and $a$'s are more direct.

The main purpose of this paper is to build a basis for  ${\cal L}_{6}$ such that each $d=6$
operator ${\cal O}_{a}$
is related to a coupling $a$,
\begin{equation}
\label{Leff2}
{\cal L}_{6} = \sum_{ a} \frac{c_a}{\Lambda^2}  \,  {\cal O}_{a} \ .
\end{equation}
Here the sum runs over a chosen set of couplings $\{ a \}$. 
In order that  the set of gauge-invariant operators $\{ {\cal O}_a \}$ is a basis, a necessary condition is that we have
as many couplings $a$ in the sum in (\ref{Leff2}) as operators $i$ in (\ref{Leff1}).
We have some freedom in choosing which couplings form such a complete set;
whenever it is possible we will  elect the couplings that are better measured. 
We may say that $\{ {\cal O}_a \}$ is a ``coupling basis".

 At the Lagrangian level, 
a  coupling $a$ corresponds to a coefficient multiplying an operator
$ \widehat{ {\cal D}}_a$, formed by  SM fields.
We say that $a$ and ${\cal O}_{a}$ are related when 
the operator $ \widehat{ {\cal D}}_a$  is contained in ${\cal O}_{a}$ when the latter is
written in the unitary gauge,
\be
\frac{c_a}{\Lambda^2}  \, {\cal O}_{a}^{unit} = 
\eta_{a} \left(  \widehat{ {\cal D}}_a +\delta {\cal D}_a \right) \ .
\label{Da}
\ee
The coefficients $\eta_{a}$ are a measure of potential
deviations from SM predictions (we choose the $\eta_a$'s adimensional).
In \eq{Da},  $\delta {\cal D}_a$ stands for a series of coupling terms that inevitably accompany $\widehat{ {\cal D}}_a$.
We define
\be
{\cal D}_a= \widehat{ {\cal D}}_a +\delta {\cal D}_a \ .
\label{Da2}
\ee
From a practical point of view, the Lagrangian in the unitary gauge to be added to the SM, which describes
 BSM effects, reads
\be
\Delta {\cal L} =  \sum_{ a}  \eta_{a}  {\cal D}_a \ .
\label{LDa}
\ee
This Lagrangian is written in the unitary gauge and is to be used at tree-level. Radiative corrections should be calculated with
the full ${\cal L}_{6}$ in \eq{Leff2}.

A crucial point of our work is that we manage to disentangle the couplings in  $\{ a \}$, i.e.,  ${\cal D}_a$ 
does not contain ${\widehat {\cal D}}_b$, for any $a$ and $b$ ($ b\neq a$)  in the set $\{ a \}$ over which
the sum in (\ref{Leff2})  extends. Then, measurement of all couplings $\{ \eta_a \}$ determines
the full Lagrangian.
Given $a$  in the chosen set $\{ a \}$ the expressions for ${\cal D}_a$ and ${\cal O}_a$
are unique, up to partial integrations, field redefinitions, etc.
Thus, the coupling $a$  generates a well-defined ``direction" in the operator space. With a bit of language abuse
we refer  to $a$, ${\cal D}_a$ and ${\cal O}_a$ as ``directions".

To reach our goals 
we proceed in two steps. First, we choose what we call the
``starting basis" and the set of couplings $\{ a \}$. The starting basis is a basis constituted by monomial operators ${\cal O}_{i}$ with the criterium that it
is close to the $\{ {\cal O}_{a} \}$ basis we are looking for.
There is no precise definition of ``close"; we simply mean that since we have some freedom in the election
of $\{ {\cal O}_{i} \}$, we will choose a basis that simplifies the algebra to be done in the second step.

This second step consists in disentangling the starting basis $\{ {\cal O}_{i} \}$, so that we obtain the coupling basis $\{ {\cal O}_{a} \}$.
To do it, we will work in the unitary gauge and build first the directions $\{ {\cal D}_{a} \}$, taking care that 
a given $ \widehat{ {\cal D}}_a$ appears only in
one of the directions, i.e., we have to define
independent linear combinations of operators  in the set $\{ {\cal O}_{i} \}$ such
that each combination contributes to one coupling  of the set $\{ a \}$ and not to  any other.
These linear combinations will give the desired $\{ {\cal O}_{a} \}$ basis.

Once this program is done, we may say that  \eq{Leff2} is  the most general $d=6$ effective Lagrangian 
taking as directions a set of  vertices that can be determined (or constrained) experimentally, with the directions
independent one from the other. We could work out this program for all operators in a complete basis,
which is formed by 59 operators in the case of one family\footnote{We do not consider $B$-violating operators.}  
\cite{Grzadkowski:2010es}.
However, we will not consider all 59 operators; we shall ignore the four-fermion operators 
and two operators involving only gluons.
Another way to define our sub-basis is by requiring that the operators contain the Higgs field,
plus two operators containing electroweak gauge bosons only.
With this, we restrict our work to the sector relevant for electroweak data from Higgs physics, 
 electroweak precision tests (EWPT) and/or triple gauge boson coupling (TGC) measurements.
These restriction leaves us with 32 operators (for one family). We will use the term
``basis" for them, because it is indeed a basis in the sector of experiments we consider, although
they form a sub-basis of the total group of 59.
 In Appendix B we carefully proof that our starting basis is complete in this sector .

Our results may be useful to understand some features of effective Lagrangians.
Perhaps the most interesting is that it allows
to see correlations among couplings in a very transparent way. For example, concerning couplings in Higgs physics,
we can easily deduce which ones have 
discovery potential for new physics, and which ones are already constrained by EWPT tests 
and/or TGC measurements.
For this reason we believe that the form of our effective Lagrangian  may guide the search for BSM in the electroweak sector.
In Section \ref{conclusion} we discuss other potential advantages of our framework.

Let us mention the relation of our paper with other studies. Most of the work done in the field of electroweak
effective Lagrangians is done using operators that are monomials. Recently, in \cite{Pomarol:2013zra},  some combinations of monomials
have been used, with the aim 
of  making the connection with observables more direct, so that one can constrain the effects
of effective Lagrangians in a more systematic and hierarchical way. We give a step further because we
disentangle the directions completely. Actually, our work is closer in spirit to the work done in \cite{Gupta:2014rxa}.
In this paper, the authors find the most general Lagrangian describing the dominant potential deviations from SM. They
do it starting from the SM fields and using a variety of arguments. 
The construction of the Lagrangian in \cite{Gupta:2014rxa} is thus bottom-up. We build the Lagrangian using $d=6$
operators so that our approach is top-down. When we consider our Lagrangian in \eq{Leff2} in the unitary gauge
and we perform a series of redefinitions of SM parameters, i.e., our \eq{LDa}, we reproduce the result in
\cite{Gupta:2014rxa}, as expected.

We organise the paper as follows.
In Section \ref{basis} we introduce the starting basis $\{ {\cal O}_i \}$
and in Section \ref{couplings} the set of couplings $\{ a \}$.
Afterwards, we calculate the CP-even and CP-odd directions in  Section \ref{even} 
and \ref{odd}, respectively. 
We write the relations among the starting basis, the coupling basis, and the directions in Section \ref{observablebasis}.
In Section \ref{conclusion} we  discuss
our results, in particular some advantages of our approach are described there.
We include two Appendices. In Appendix A we write the SM Lagrangian in order to establish some notation.
In Appendix B we proof the equivalence of our basis of operators and the basis in \cite{Grzadkowski:2010es}.

\section{The Starting Basis}
\label{basis}

In this Section we choose a monomial basis $\{ {\cal O}_i  \}$ from which we can get 
 the coupling basis $\{ {\cal O}_{a} \}$ in a simple way. 
 Our election is closely related to the set of couplings which we present in the next Section.
 For the operators we adopt the notation in \cite{Giudice:2007fh} and \cite{Elias-Miro:2013mua}.
Further discussion of our starting basis is done in Appendix B

We separately consider the CP-even and CP-odd cases.

\subsection{CP-even Basis}

We first introduce operators which can only be tested in Higgs physics, i.e., 
operators such that do not lead to any physical effect
 when the physical Higgs field is set equal to zero.
For one family, the total number of such operators is eight  \cite{Elias-Miro:2013mua}. 
There are different possible choices; our election is motivated by simplicity in achieving our goals:
We choose eight operators most directly related to vertices which
are or will be extracted from experiment.

We have five bosonic operators 
\begin{eqnarray}
{\cal O}_r=|\H |^2 |D_\mu \H |^2 & , &
{\cal O}_6=\lambda |\H|^6   \ , \nonumber \\
{\cal O}_{BB} = {g}^{\prime 2} |\H|^2 B_{\mu\nu}B^{\mu\nu} & , &
{\cal O}_{WW}={g}^{2} |\H|^2 \cW^a_{\mu\nu} \cW^{a\, \mu\nu}  \, , \nonumber \\
  {\cal O}_{GG}=g_s^2 |\H|^2 \cG_{\mu\nu}^A \cG^{A\mu\nu} & . & 
\label{first6dimA}
\end{eqnarray}
Here $\Phi$ is the complex Higgs doublet, and $\cG_{\mu\nu}^A$ is the $SU(3)_c$ gluon field strength; for
other notation see Appendix A.
We also have three operators involving fermions
\begin{eqnarray}
{\cal O}_{y_u}   = y_u |\H|^2    \bar Q_L \widetilde{\H} u_R  & , &
{\cal O}_{y_d}   = y_d |\H|^2    \bar Q_L {\H} d_R  \, , \nonumber \\
{\cal O}_{y_e}   = y_e |\H|^2    \bar L_L {\H} e_R  & , &
\label{first6dimB}
\end{eqnarray}
where  $\widetilde{\H}=i\sigma_2 \H^*$. 
Here and in the rest of the paper we restrict our analysis to one family.

Indeed, we see that in vacuum, $\H^T=(0,v/\sqrt{2})$, 
the operators in \eq{first6dimA} and in  \eq{first6dimB}
 lead to innocuous redefinitions of SM Lagrangian parameters.
We shall refer to this group of eight operators as ``Higgs-only" operators.

There is a second class of operators which can be tested in $h$-physics as well as in EWPT and TGC physics. 
There are two bosonic operators
\bea
{\cal O}_B - {\cal O}_W &=& \frac{ig'}{2}( \H^\dagger  \lra {D^\mu} \H )\partial^\nu  B_{\mu \nu}   \ - \ 
 \frac{ig}{2}( \H^\dagger  \sigma^a \lra {D^\mu} \H )D^\nu  \cW_{\mu \nu}^a  \nn \\
{\cal O}_{HB}  &=& i g' (D^\mu \H)^\dagger(D^\nu \H)B_{\mu\nu} \, ,
\label{second6dimA}
\eea
where $\H^\dagger {\lra { D_\mu}} \H\equiv \H^\dagger D_\mu \H - (D_\mu \H)^\dagger \H $.
The first operator in \eq{second6dimA} is the difference of two monomials, rather than a single one.
 This is not a problem, after all we are going to combine all these operators in the
starting basis to get the coupling basis. The reason why we choose such a combination will become
clear below, in Eqs. \eq{OW} and \eq{OB}.

In this second class, we also have fermionic operators. There are seven products of Higgs and fermion currents
\bea
{\cal O}^u_{R}  = 
(i \H^\dagger {\lra { D_\mu}} \H)( \bar u_R\gamma^\mu u_R) & , & 
{\cal O}^d_{R} =
(i \H^\dagger {\lra { D_\mu}} \H)( \bar d_R\gamma^\mu d_R)\, ,
\nonumber\\
{\cal O}^e_{R}  = 
(i \H^\dagger {\lra { D_\mu}} \H)( \bar e_R\gamma^\mu e_R) & , & 
\nonumber\\
{\cal O}^q_{L} = 
(i \H^\dagger {\lra { D_\mu}} \H)( \bar Q_L\gamma^\mu Q_L) & , & 
{\cal O}^l_{L} =
(i \H^\dagger {\lra { D_\mu}} \H)( \bar L_L\gamma^\mu L_L)\, ,
\nonumber\\
{\cal O}_{L}^{(3)\, q} = 
(i \H^\dagger \sigma^a {\lra { D_\mu}} \H)( \bar Q_L\gamma^\mu\sigma^a Q_L) & , & 
{\cal O}_{L}^{(3)\, l} =
(i \H^\dagger \sigma^a {\lra { D_\mu}} \H)( \bar L_L\gamma^\mu\sigma^a L_L)\,  .
\label{second6dimB}
\eea
In addition, there are eight dipole operators
\begin{eqnarray}
{\cal O}^u_{DB}  = y_u g' \,  (\bar Q_L \sigma^{\mu \nu} u_R\, \widetilde \H ) \, B_{\mu \nu} & , & 
{\cal O}^u_{DW} = y_u g \, (\bar Q_L \sigma^{\mu \nu} u_R\,  \sigma^a \widetilde \H  ) \,  \cW^a_{\mu \nu}\ , \nonumber\\
{\cal O}^d_{DB}  = y_d g' \, (\bar Q_L \sigma^{\mu \nu} d_R\,  \H  ) \,  B_{\mu \nu} & , & 
{\cal O}^d_{DW} = y_d g \, (\bar Q_L \sigma^{\mu \nu} d_R\,  \sigma^a   \H  ) \,  \cW^a_{\mu \nu} \ , \nonumber\\
{\cal O}^e_{DB}  = y_e g' \, (\bar Q_L \sigma^{\mu \nu} e_R\,  \H  ) \, B_{\mu \nu} & , & 
{\cal O}^e_{DW} = y_e g (\bar Q_L \sigma^{\mu \nu} e_R\,  \sigma^a   \H ) \,  \cW^a_{\mu \nu}\ , \nonumber\\
{\cal O}^u_{DG} = y_u g_s \,
 \bar Q_L \sigma^{\mu \nu} T^A u_R\, \widetilde \H \, \cG^A_{\mu \nu} & , &
 {\cal O}_{DG}^d= y_d g_s \,
 \bar Q_L \sigma^{\mu \nu} T^A d_R\,  \H  \, \cG^A_{\mu \nu} \ , 
\label{second6dimC}
\end{eqnarray}
and, finally, an operator involving  $u_R$ and $d_R$
\be
{\cal O}_{R}^{ud} =
y_u y_d(i \widetilde \H^\dagger {\lra { D_\mu}} \H)( \bar u_R\gamma^\mu d_R) \ .
\label{second6dimD}
\ee
We refer to the whole class of 18 operators \eq{second6dimA}-\eq{second6dimD} as ``Higgs-EWPT-TGC" operators.
Some of these operators come with its hermitian conjugate.

There is a third group which actually consists of a single operator 
that can be only constrained using TGC measurements
\be
{\cal O}_{3W}= \frac{g}{3!} \, \epsilon^{abc} \, \cW^{a\, \nu}_{\mu}\,  \cW^{b}_{\nu\rho} \, \cW^{c\, \rho\mu}  \ .
\label{3W}
\ee

Counting the operators in \eq{first6dimA}-\eq{3W} gives a total number
\be
N_{CP+} = 27  
\label{NCPeven}
\ee
of  CP-even operators in the starting basis.
 As we said, we restrict to one fermion family.

\subsection{CP-odd Basis}

We now address our interest to CP-violating operators. Actually, many operators of a complete basis,
 when multiplied by a complex coefficient have a CP-odd part.  
A classical example are the dipole operators in \eq{second6dimC}; 
 if we allow for complex coefficients we have magnetic and (CP-odd) electric dipoles.

Apart from the possibility of complex coefficients, there are operators that are genuinely CP-odd.
Three of them are the CP-odd siblings of $ {\cal O}_{BB},  {\cal O}_{WW},  {\cal O}_{GG}$ in \eq{first6dimA},
\begin{eqnarray}
 {\cal O}_{B \widetilde{B}} = {g}^{\prime 2} |\H|^2 B_{\mu\nu} \widetilde{B}^{\mu\nu} & , &
{\cal O}_{W \widetilde{W}}={g}^{2} |\H|^2 \cW^a_{\mu\nu} \widetilde{\cW}^{a\, \mu\nu}  \, , \nonumber \\
  {\cal O}_{G \widetilde{G}}=g_s^2 |\H|^2 \cG_{\mu\nu}^A \widetilde{\cG}^{A\mu\nu} & , & 
  \label{Oodd1}
\end{eqnarray}
with  $\widetilde V^{\mu\nu}=\epsilon^{\mu\nu\rho\sigma} V_{\rho\sigma}/2$.
The other two are the CP-odd versions of  ${\cal O}_{HB} $ in \eq{second6dimA} and ${\cal O}_{3W}$ in \eq{3W},
\be
{\cal O}_{H\widetilde B}=ig'(D^\mu \H)^\dagger(D^\nu \H)\widetilde B_{\mu\nu} \ \ , \ \
{\cal O}_{3\widetilde W}=\frac{1}{3!} \ g\, \epsilon^{abc} \  \cW^{a\, \nu}_{\mu} \cW^{b}_{\nu\rho} \widetilde \cW^{c\, \rho\mu} \ .
  \label{Oodd2}
\ee
We have a total of
\be
N_{CP-} = 5  
\label{NCPodd}
\ee
 CP-odd operators in the starting basis.

\section{The couplings}
\label{couplings}

In this Section we list the set of couplings $\{ a \}$ which generate the basis $\{ {\cal O}_a \}$.
The set has $N_{CP+} = 27$ CP-even and $N_{CP-} = 5$ CP-odd couplings.

\subsection{CP-even Couplings}

We start with eight couplings of the physical Higgs $h$:
 to fermions (for one family), the cubic self-coupling, the custodial preserving
combination of $hWW$ and $hZZ$ (custodial-violating combinations will appear in some of the $\delta {\cal D}_{a}$),
and finally to gluons, photons and $\gamma Z$,
\bea
{\widehat {\cal D}}_{hff} &=& h  \, \left[   \bar f_L f_R + {\rm h.c.} \right] \ , \ f=u,d,e \ ,  \ \ \ 
{\widehat {\cal D}}_{h3} =v h^3 \ , \nn \\
{\widehat {\cal D}}_{h(VV)_c} &=&  vh \ \left(  W^{+\mu} W^-_\mu + \frac{1}{2 \ctw^2}  Z^\mu Z_\mu \right)  \ , \nn \\
{\widehat {\cal D}}_{hgg}  &=&  \frac{h}{v} \,  G_{\mu\nu}^A G^{A\mu\nu} \ , \ \ \
{\widehat {\cal D}}_{h\gamma\gamma} = \frac{h}{v} \,  A^{\mu \nu} A_{\mu \nu}   \ , \ \ \
{\widehat {\cal D}}_{h\gamma Z} = \frac{h}{v} \,  A^{\mu \nu} Z_{\mu \nu} \ ,
\label{couplingsOnlyHiggs}
\eea
where 
\be
V_{\mu \nu} = \partial_\mu V_\nu - \partial_\nu V_\mu \ ,
\label{abelian}
\ee
is the abelian part of field strength.
 Some of these $h$-couplings are being extracted from measurements at LHC and some will be hopefully measured in the future.

We  continue listing the set $\{ a \}$  with three  couplings measured in processes involving TGC. We define
$a= g1Z, \kappa\gamma$ as
\bea
{\widehat {\cal D}}_{g1Z}  &=& i \ctw g \left[ Z^{\mu}  \left( W^{+ \, \nu}     W^-_{\mu\nu}  - W^{- \, \nu}  W^+_{\mu\nu} \right) 
+ Z^{\mu \nu} W^+_\mu W^-_\nu \right]
\nn \ , \\
{\widehat {\cal D}}_{\kappa \gamma} &=&  i e    W^+_\mu  W^-_\nu \left( A^{\mu\nu} - \ttw Z^{\mu\nu} \right) \ ,
\label{g1Zkappagamma}
\eea
and $a= \lambda V$ as
\be
\widehat   {\cal D}_{\lambda V} =  \frac{i}{m^2_W}  \left(g  \ctw \, Z^{\mu \nu} + e \, A^{\mu \nu} \right   )   
W^{- \rho}_{\nu}       W^{+}_{\rho \mu} \ .
\label{couplings lambda}
\ee

These couplings  appear in the general expression for the CP-even TGC vertex
\cite{Hagiwara:1986vm}. Restricting to terms generated by $d=6$ operators, we have
\bea
\delta {\cal L}_{3V, CP+} &=&  i \ctw g 
\left[ 
\delta g_1^Z \, Z^{\mu}  \left( W^{+ \, \nu}     W^-_{\mu\nu} 
  - W^{- \, \nu}        W^+_{\mu\nu} \right)  \right]  \nonumber    \\ 
  &+ &
   i g \sum_{V=\gamma, Z}    \, c_V \, 
\left[  \delta \kappa_V \,      V^{\mu\nu} W^+_\mu  W^-_\nu
  +  \ 
  \frac{\lambda_V}{m^2_W}      V^{\mu\nu}       W^{- \rho}_{\nu}       W^{+}_{\rho \mu}
 \right]  \ ,
\label{V3}
\eea
where $c_\gamma=\stw, c_Z=\ctw$,
and  we follow the notation shown in \eq{abelian} for the photon $A$, the $Z$-boson and the $W^\pm$-bosons.

The Lagrangian ${\cal L}_6$ implies some relations among the TGC parameters in \eq{V3},
 coming from custodial invariance preserved by the relevant $d=6$ operators and 
 electric charge conservation \cite{Hagiwara:1993ck}
\be
\delta \kappa_Z= \delta g_1^Z - \ttw^2 \delta \kappa_\gamma \  , \ \ \
 \lambda_Z=\lambda_\gamma \  .
 \label{lambdagammaZ}
 \ee
 The five parameters in \eq{V3} together with the two equations \eq{lambdagammaZ} give
 three independent parameters.
 We take them to be $ \delta g_1^Z$, $ \delta \kappa_\gamma$ and 
$\lambda_V$ in the set $\{ a \}$.
The first operator in \eq{g1Zkappagamma} corresponds to $\delta g_1^Z=\delta \kappa_Z$ 
and the second to $ \delta \kappa_\gamma= \ttw^{-2} \delta \kappa_Z$,  while other couplings are
set equal to zero.
The relation $\lambda_Z=\lambda_\gamma$  in \eq{lambdagammaZ} leads
us to define the coupling the way we do it in \eq{couplings lambda}.
The deviation of the TGC couplings from the SM prediction was constrained by
 LEP-2, and is supposed to be further investigated by LHC.

We continue with seven couplings that are very well measured at the $Z$-pole,
\be
\widehat {\cal D}_{Z eR} =  Z_\mu \, \bar e_R \gamma^\mu e_R  \ , \ \
\widehat {\cal D}_{Z uR}  =  Z_\mu  \,  \bar u_R \gamma^\mu u_R  \ , \ \ 
\widehat {\cal D}_{Z dR}  =   Z_\mu  \,  \bar d_R \gamma^\mu d_R   \ ,
\label{dirZff1}
\ee
and
\begin{eqnarray} 
\widehat  {\cal D}_{Z eL}  =   Z_\mu  \, \bar e_L \gamma^\mu e_L &, & \  \
\widehat {\cal D}_{Z uL}  =    Z_\mu  \, \bar u_L \gamma^\mu u_L \ , \
 \nonumber  \\
\widehat   {\cal D}_{Z dL}  =  Z_\mu  \,  \bar d_L \gamma^\mu d_L 
& , & \  \  \widehat   {\cal D}_{Z \nu L} = Z_\mu  \,  \bar \nu_L \gamma^\mu \nu_L \ . \
\label{dirZff2}
\end{eqnarray}
These seven couplings are tightly bounded from $Z$-pole measurements.
We do not use the $S$ parameter \cite{Peskin:1990zt} because it would be
redundant. Indeed, $S$ is equivalent to a universal part of the seven couplings
\eq{dirZff1} and \eq{dirZff2}.

In our list there also are eight dipole couplings, defined as
\be
\widehat {\cal D}_{DGq}  =  \frac{1}{v}\, \bar q_L \sigma^{\mu  \nu} T^A q_R \, G^A_{\mu \nu}+ {\rm h.c.} 
\ , \ \ \ \
\widehat {\cal D}_{DVf} = \frac{1}{v} \,  \bar f_L \sigma^{\mu \nu} f_R V_{\mu \nu}  +  {\rm h.c.} \ ,
\label{couplings Dipole}
\ee
with $q=u,d$  and $V=A,Z$, $f=u,d,e$. See Ref. \cite{Contino} for  updated constraints on dipole couplings.
Finally, we also include 
\be
\widehat   {\cal D}_{uRdR} =  \bar u_R \gamma^\mu d_R W^+_\mu + {\rm h.c.} \  \ .
\label{couplings Wurdr}
\ee
The total number of couplings in Eqs. \eq{couplingsOnlyHiggs}, \eq{g1Zkappagamma},
\eq{couplings lambda}, and \eq{dirZff1}-\eq{couplings Wurdr} is 
$N_{CP+} = 27$.

\subsection{CP-odd Couplings}

In the list of CP-odd couplings 
we have  the three vertices $h$-gluon-gluon, $h$-photon-photon, and $h$-photon-$Z$,
\be
\widehat   {\cal D}_{ hg \widetilde g } =  \frac{h}{v} \, G_{\mu\nu}^A \widetilde G^{A\mu\nu} 
\ , \ \ \ \
\widehat  {\cal D}_{h\gamma \widetilde \gamma} = \frac{h}{v}  \, A_{\mu \nu}  \widetilde A^{\mu \nu} 
\ , \ \ \ \
\widehat  {\cal D}_{h\gamma   \widetilde  Z} = \frac{h}{v}  \, A_{\mu \nu}  \widetilde Z^{\mu \nu}  \ ,
\label{coupling CPodd onlyH}
\ee
and two more appearing in the expression for the CP-odd TGC vertex
 \cite{Hagiwara:1986vm}
\be
\delta {\cal L}_{3V,CP-} =  i g \sum_{V=\gamma, Z}  \, c_V \, 
\left[     \delta \widetilde{\kappa}_V \,   \widetilde   V^{\mu\nu} W^+_\mu  W^-_\nu 
+ \frac{\widetilde \lambda_V}{m^2_W}      \widetilde V^{\mu\nu}      W^{- \rho}_{\nu}       W^{+}_{\rho \mu} \right]
 \ ,
\label{V3odd}
\ee
where $c_\gamma=\stw, c_Z=\ctw$. We have restricted to parameters induced by $d=6$ operators.
In addition, one has that
\be
\delta \widetilde{\kappa}_Z = - \ttw^2 \, \delta \widetilde{\kappa}_\gamma \ , \ \
\widetilde \lambda_\gamma = \widetilde \lambda_Z \ .
\ee
We take as couplings in the set $\{ a \}$ the following two:
\bea
\widehat  {\cal D}_{ \widetilde{\kappa \gamma} }  &=&  i e\,  W^+_\mu W^-_\nu   \left( \widetilde  A^{\mu \nu} 
- \ttw \widetilde Z^{\mu \nu} \right) \ , \nn \\
\widehat  {\cal D}_{\widetilde {\lambda V}} &=&  \frac{i }{M^2_W}  
\left(g  \ctw \widetilde Z^{\mu \nu} + e \, \widetilde A^{\mu \nu} \right)   W^{- \rho}_{\nu} \, W^+_{ \rho \mu}   \ .
\label{coupling CPodd V3}
\eea
There are $N_{CP-} = 5$ CP-odd couplings, as expected.
 
 \subsection{From Couplings to Directions}
 
 In the next Section we build the set of directions $\{ {\cal D}_a \}$. The material we have is on the one hand the starting basis
 operators in Section \ref{basis} and on the other hand the couplings we have introduced in the present Section.
 The attentive reader has noticed a parallelism among groups of operators in Section
 \ref{basis} and groups of couplings in Section  \ref{couplings}.
   For example, the eight Higgs-only operators  in \eq{first6dimA} and \eq{first6dimB}
 are related to the eight Higgs couplings in \eq{couplingsOnlyHiggs}.
 This of course has a reason. 
 The number has to be the same, if not we would have either a redundancy or missing elements in the basis.
However,
 not all possibilities are allowed.
  It is not possible to change the elected couplings 
 arbitrarily. For instance one would be tempted to dispose of the $h^3$ coupling, and to elect another one that will
 presumably be measured before, for example a non-custodial $hWW$-$hZZ$ coupling. This is not allowed, because
in any $d=6$ basis one has one operator giving essentially $h^3$, so that we cannot ignore it.
 There are allowed possibilities to choose other couplings instead
 of the election \eq{couplingsOnlyHiggs}. However, this new election would mean to introduce couplings that are less
 well measured, and this is why we stick to \eq{couplingsOnlyHiggs}. 
 
Apart from the Higgs-only sector,  the parallelism we have  alluded is obvious in the correspondence 
\eq{second6dimB}$\leftrightarrow$\eq{dirZff1}-\eq{dirZff2}, \eq{second6dimC}$\leftrightarrow$\eq{couplings Dipole}, 
 \eq{second6dimD}$\leftrightarrow$\eq{couplings Wurdr}, and \eq{3W}$\leftrightarrow$\eq{couplings lambda}. We will also see that there is a correspondence
\eq{second6dimA}$\leftrightarrow$\eq{g1Zkappagamma}. 
Finally, for the CP-odd sector the correspondence reads \eq{Oodd1}$\leftrightarrow$\eq{coupling CPodd onlyH},
and \eq{Oodd2}$\leftrightarrow$\eq{coupling CPodd V3}.

\section{The CP-even Directions}
\label{even}

In this Section we calculate the directions ${\cal D}_a$, in the case that ${\cal L}_{6}$
respects CP. We use the notation shown in Appendix A, with $H/\sqrt{2}$ 
being the neutral part of the Higgs doublet, and
$H=v+h$, with $h$ the physical Higgs field and $v \simeq 243$ GeV as given in \eq{vDef}.

\subsection{Higgs-only Sector}

Now we use the eight operators in \eq{first6dimA} and  \eq{first6dimB} 
 to find the eight directions corresponding to the couplings \eq{couplingsOnlyHiggs}.

\subsubsection{  $ \mathbf{ a=h\, f \, f} $ }

The three Higgs-fermion-fermion directions $a=hff$ can be simply obtained from
\begin{equation}
\label{Oy} 
{\cal O}_{hff} = {\cal O}_{y_f} \rightarrow H^3 \, \left(  \bar f_L f_R + {\rm h.c.} \right) \ ,
\end{equation}
with $f=u,d,e$.
Here and in the following we denote by an arrow the expression of the $d=6$ operators in the unitary gauge,
up to signs, constant factors, and/or coupling constants. 

 We write the factor $H^3$ in  \eq{Oy}  as
\begin{equation}
\label{}
H^3 = H\, v^2 + H\,  ( H^2 - v^2 ) \ .
\end{equation}
In this expression the part $H v^2$ leads to a term proportional to the Yukawa term $y_f H \bar f f$ in the SM Lagrangian,
and it can reabsorbed by redefining $y_f$.
Expanding the  part $H( H^2 - v^2 )$  leads to
\begin{equation}
\label{hff1}
{\cal D}_{hff} = ( h \, P_2 )  \, \left[   \bar f_L f_R + {\rm h.c.} \right]  \ ,
\end{equation}
with
\be
P_2= 1 +  \frac{3 h}{2  v} +   \frac{h^2}{2 v^2}  \ .
\label{P2}
\ee
Here and in the following $P_n, \dots$ are order-$n$ polynomials in $h$.
They all start with an independent term equal to one,
so that higher $h$-powers  lead to vertices with additional $h$'s in the vertex. 
For example, in \eq{hff1} we see that setting $P_2=1$ gives us the desired operator 
$\widehat{ {\cal D}}_{hff}=h (\bar f_L f_R + {\rm h.c})$, which
defines the coupling. As expected, accompanying the operator $\widehat{ {\cal D}}_{hff}$ there
are other terms $\delta {\cal D}_{hff}$, see  \eq{Da} and \eq{Da2}. 
One can easily  generalise this result to the case of three families.
To obtain ${\cal D}_{hff}$ from ${\cal O}^{unit}_{hff}$
 we have to make an unobservable redefinition of the Yukawa couplings $y_f$.

\subsubsection{ $\mathbf{ a=h\, 3}$} 

For the triple Higgs vertex $h^3$, which hopefully will be measured in the future, we only
need ${\cal O}_6$,
\begin{equation}
\label{O6}
{\cal O}_{h3} = {\cal O}_6 \rightarrow H^6 \ ,
\end{equation}
which can be written as
\begin{equation}
\label{O6_1}
 H^6 \sim \left( H^2 - v^2 \right)^3   \ .
\end{equation}
Indeed, when expanding the cube power in the rhs, the terms $H^4$ and $H^2$ can be absorbed in parameters of the $V(H)$ potential,
see \eq{VH}. The symbol $\sim$ will be used to relate terms in the Lagrangian that are equivalent up to redefinitions and/or partial
integrations.

From \eq{O6_1} we get the expression
\begin{equation}
\label{h3}
{\cal D}_{h3} = v h^3 P_1^3 \ ,
\end{equation}
with
\be
P_1= 1 +    \frac{h}{2  v}  \ .
\label{P1}
\ee

To obtain ${\cal D}_{h3}$ from ${\cal O}^{unit}_{h3}$
 we have to make an unobservable redefinition of the  $V(H)$ parameters.
 
\subsubsection{$\mathbf{ a=h\, (V V)_c}$}
\label{hVVsubS}

Next we consider the coupling of $h$ to $WW$ and $ZZ$ in \eq{couplingsOnlyHiggs}.
The relevant operator in the starting basis is ${\cal O}_r$.
It can written as
\be
{\cal O}_r  \sim   \left( |\H |^2 - \frac{v^2}{2} \right) |D_\mu \H |^2 \ ,
\label{Or}
\ee
because the part proportional to $ v^2$ adds to the kinetic $\H$-term and it can be reabsorbed
by redefining $\H$ and bare parameters in the SM Lagrangian.

\Eq{Or}, in the unitary gauge, gives
\be
\frac{1}{4} (H^2 - v^2) \left( \partial^\mu H \partial_\mu H + \frac{g^2}{2}  H^2 W^{+\mu} W^-_\mu + \frac{g^2}{4\ctw^2}  H^2 Z^\mu Z_\mu \right) \ .
\ee
We  get the desired term $h(VV)_c$, but to get rid of the term  $(\partial^\mu H)^2$ we  integrate by parts,
 \be
 (H^2 - v^2) \left( \partial^\mu H \partial_\mu H \right) \sim - \left( v + \frac{h}{3} \right)\,  h^2 \, \partial^\mu \partial_\mu  H
 \ee
and use  the EoM in \Eq{eomH}  for $\partial^\mu \partial_\mu  H$.
The resulting expression contains  a $h^3$-term. Since we wish the directions to be independent one from the
other, we have to subtract a term proportional to \eq{h3}. The end result for
 the $h(VV)_c$ direction is given by
\be
 {\cal D}_{h(VV)_c}  =  v (h P_3)  \left[  W^{+\mu} W^-_\mu + \frac{1}{2 \ctw^2}  Z^\mu Z_\mu \right]
 +  \frac{m_f}{4 m_W^2} (h^2  Q_1) {\bar f} f
+ \frac{m_h^2}{12 m_W^2} (h^4 Q_2)  \ ,
\label{OhVV}
\ee
where there is a sum over all fermions $f$, and we have defined
\bea
P_3 &=& 1 + \frac{2 h}{v} +   \frac{4 h^2}{3 v^2} +   \frac{h^3}{3 v^3}   \ ,  \nn \\
Q_1 &=& 1 +   \frac{h}{3 v}    \ ,    \nn \\
Q_2 &=& 1 +  \frac{3 h}{4 v} +  \frac{h^2}{8 v^2}  \ .
\label{Q1}
\eea

We notice the appearance of the  custodial-preserving combination 
 \be
{\widehat {\cal D}}_{h(VV)_c}=  vh \ \left(  W^{+\mu} W^-_\mu + \frac{1}{2 \ctw^2}  Z^\mu Z_\mu \right)    \ .
\label{ChVV}
\ee
as the coupling $a=h(VV)_c$ in the basis set $\{ a\}$,
as we anticipated in \eq{couplingsOnlyHiggs}.

At the $d=6$ operator level, this direction is obtained from the combination of two
operators in the starting basis: ${\cal O}_{r}$ and ${\cal O}_{6}$. In order that the latter cancels
de cubic Higgs terms, the precise combination is 
\be
{\cal O}_{h(VV)_c}={\cal O}_{r} - \frac{1}{2} {\cal O}_{6} \ .
\label{hVVr6}
\ee
 To get 
${\cal D}_{h(VV)_c}$ from ${\cal O}_{h(VV)_c}$, one has to set the unitary gauge and redefine SM parameters,
specifically the Yukawa couplings and $V(H)$ parameters.

\subsubsection{$\mathbf{ a=g\,g}$}
\label{ssgg}

To determine the direction Higgs-gluon-gluon, 
 in \eq{couplingsOnlyHiggs}, we only need $ {\cal O}_{GG}$,
\begin{equation}
\label{OggA}
{\cal O}_{hgg}= {\cal O}_{GG}  \rightarrow H^2 \, \cG_{\mu\nu}^A \cG^{A\mu\nu}    \ .
\end{equation}
We decompose the factor $H^2$ in \eq{OggA} as
\be
H^2 = v^2 + ( H^2 - v^2 )    \ .
\label{decomposition}
\ee
The term $v^2 \cG_{\mu\nu}^A \cG^{A\mu\nu}$ can be ignored because it is
proportional to the kinetic term for the gluon field, and thus can be reabsorbed
in the coupling $g_s$. Therefore, 
\begin{equation}
\label{Ogg}
{\cal D}_{hgg}  =  \left( \frac{h}{v} P_1 \right) \cG_{\mu\nu}^A \cG^{A\mu\nu}    \ .
\end{equation}
Notice that we get \eq{Ogg} from $ {\cal O}_{GG}$ when redefining  $g_s$.

\subsubsection{$\mathbf{ a=h \, \gamma \, \gamma , \ h \, \gamma\,  Z}$}
\label{ssgggZ}

The directions  ${\cal O}_{h\gamma\gamma}$ and ${\cal O}_{h\gamma Z}$ correspond to combinations of ${\cal O}_{BB}$ and ${\cal O}_{WW}$.
We have
\bea
{\cal O}_{BB}  & \rightarrow & H^2  \, B_{\mu\nu} B^{ \mu\nu} \sim (H^2 - v^2)  \, B_{\mu\nu} B^{ \mu\nu}    \ ,  \nn \\
{\cal O}_{WW}  & \rightarrow & H^2 \, \cW^a_{\mu\nu} \cW^{a\, \mu\nu} \sim (H^2 - v^2)  \, \cW^a_{\mu\nu} \cW^{a\, \mu\nu}    \ .
\label{hgghgZ}
\eea
where we obtain the two rhs parts using \eq{decomposition} and reabsorbing the corresponding $v^2$-part
in $g'$ and $g$, as we have done above for $a=hgg$.
We can form two linear combinations of the rhs parts in \eq{hgghgZ}, one which cancels the vertex $h\gamma Z$
and the other which cancels $h\gamma \gamma$. The first combination defines the direction $a= h\gamma \gamma $
\be
{\cal D}_{h\gamma\gamma} = \left( \frac{h}{v} P_1 \right)  \left[ \cA_{\mu \nu} \cA^{\mu \nu} + \cZ_{\mu \nu} \cZ^{\mu \nu} + 
2\, \cW^+_{\mu \nu} \cW^{- \, \mu \nu}   \right]    \ ,  
\ee
and the second defines the direction $a=h \gamma Z$
\be
{\cal D}_{h\gamma Z} =  \left( \frac{h}{v} P_1 \right)  \left[ \cA_{\mu \nu} \cZ^{\mu \nu} + \frac{\ctw^2 - \stw^2}{2 \ctw \stw} \cZ_{\mu \nu} \cZ^{\mu \nu} + 
\frac{\ctw}{\stw} \, \cW^+_{\mu \nu} \cW^{- \, \mu \nu}   \right]    \ .
\label{DhgammaZ}
\ee
 The non-abelian parts of the field strengths $\cA_{\mu \nu}, \cZ_{\mu \nu}, \cW^{\pm}_{\mu \nu}$ are shown in \eq{amunu}.
  
In these directions, we get vertices  of the form $h  V_{\mu \nu} V^{\mu \nu}$,
$V=Z,W$. We have defined the direction $h(VV)_c$ in terms of couplings $h  V_{\mu } V^{\mu}$. These two types of couplings 
can be distinguished experimentally, so that  $h \gamma Z$,   $h \gamma \gamma$ and $h(VV)_c$ are independent directions.
The terms $h  V_{\mu \nu } V^{\mu \nu}$ in \eq{DhgammaZ} induce custodial breaking contributions to $h \to V V^{(*)}$, where
$V^{(*)}$ is an off-mass-shell $Z$ or $W$.
Finally, notice that get the ${\cal D}_a$ from the corresponding ${\cal O}^{unit}_a$ for $a=h\gamma\gamma, h\gamma  Z$,
we need to redefine the couplings $g,g'$.

\subsection{Higgs-EWPT-TGC Sector}

Let us now turn our attention to operators which induce changes in $h$-physics, 
EWPT, and TGC. 
We first discuss
 ${\cal O}_B - {\cal O}_W$ and ${\cal O}_{HB}$
in \eq{second6dimA}, which are related to the couplings
 in \eq{g1Zkappagamma}, and afterwards we will move to directions with fermions.

\subsubsection{$\mathbf{ a=g1Z}$}

Consider ${\cal O}_W$,
\be
{\cal O}_W = \frac{ig}{2}( \H^\dagger  \sigma^a \lra {D^\mu} \H )D^\nu  \cW_{\mu \nu}^a 
\, \rightarrow \, H^2 (g W^{a \mu} - g' \delta^{a3} B^\mu)  D^\nu  \cW_{\mu \nu}^a    \ .
\label{OW}
\ee
In the rhs we see there is a contribution $v^2 (g/\ctw) Z^\mu g \epsilon^{3bc} W^{b\nu} W^c_{\mu\nu}$, i.e.,
a contribution to $\delta g_1^Z $ . However, there is a contribution
$-v^2 g' B^\mu \partial^\nu W^3_{\mu\nu}$, i.e. a
 kinetic $W^3$-$B$ mixing. This contributes
 to the $S$-parameter \cite{Peskin:1990zt}, 
equivalent to an universal modification of the $Zff$ vertices.
Since we wish to keep all the $Zff$ directions -see \eq{dirZff1}-
we should subtract the contribution of ${\cal O}_W $ to $S$.
 The simplest way to proceed is to consider 
the operator \cite{Giudice:2007fh}
\be
{\cal O}_B  = \frac{ig'}{2}( \H^\dagger  \lra {D^\mu} \H ) \partial^\nu  B_{\mu \nu}  
\, \rightarrow \, H^2 Z^\mu  \partial^\nu  B_{\mu \nu}    \ ,  
\label{OB} 
\ee
which contributes to $S$ but not to $\delta g_1^Z $.
The right combination to get rid of $S$ is 
the difference ${\cal O}_B  - {\cal O}_W$ \cite{Giudice:2007fh}. This is the reason we started in \eq{second6dimA}
with such a combination.

To determine the direction corresponding to  $\delta g_1^Z$ 
we decompose the factor $H^2$ in \eq{OW} and \eq{OB} as  in \eq{decomposition}. The term
with $v^2$, after partial integration and $g,g'$ redefinitions contains the term $\delta g_1^Z$,
accompanied by quartic gauge couplings. Concerning the term with $H^2 - v^2$ we use
the EoM for the gauge fields  \eq{eomV}. The result is
\begin{eqnarray}
&&    i \ctw g  \left[ Z^\mu ( W^{+\nu} \cW^-_{\mu \nu } - {\rm h.c.}) + \cZ^{\mu \nu} W^+_\mu W^-_\nu \right] \nonumber \\
&& - \, \frac{1}{v^2} \left(   H^2 - v^2 \right) \left[  \frac{g^2 - g'^2}{4}  H^2 Z^\mu Z_\mu + \frac{g^2 \ctw^2}{2} H^2 W^{+\mu} W^-_\mu  \nonumber \right. \\
&&  \left.
+  ( g'  \stw  J^\mu_Y  + g \ctw  J^{ 3 \mu}) Z_\mu  +  \frac{\ctw^2 g }{\sqrt 2} ( J_W^\mu   W^{+}_\mu + {\rm h.c} ) \right]     \ .
\label{Og1Z1}
\end{eqnarray}
The fermionic currents are defined in \eq{currents1} and \eq{currents2}.

Notice the appearence of the vertices $hW_\mu W^\mu$ and $hZ_\mu Z^\mu$ in \eq{Og1Z1}. 
We should add to \eq{Og1Z1} a term proportional to \eq{OhVV} so that $ {\widehat {\cal D}}_{h(VV)_c} $
as given in \eq{ChVV} cancels out, i.e., $a=h(VV)_c$ and $a=g1Z$ are independent one from each other.
This will generate the presence of 
fermionic terms coupling to several Higgs and Higgs self-interactions in the direction $a=g1Z$,
\begin{eqnarray}
 {\cal D}_{g1Z}  &=&    i  g \ctw  \left[ Z^\mu ( W^{+\nu} \cW^-_{\mu \nu } - {\rm h.c.}) + \cZ^{\mu \nu} W^+_\mu W^-_\nu \right] \nonumber \\
& - &   \frac{ g^2 \ctw^2}{2}   \left( \frac{h}{v} S_2 \right) \, {\widehat{\cal D}}_{h(VV)_c}
+ \frac{g'^2}{2}   \left( h Q_3 \right) \, v \, Z^\mu Z_\mu
 \nonumber \\
&+&  g^2   \ctw^2 \left[     \frac{m_f}{4 m_W^2} (h^2  Q_1) {\bar f} f
+ \frac{m_h^2}{12 m_W^2} (h^4 Q_2) \right]
\nonumber \\
& - &   2 \left( \frac{h}{v} P_1 \right) \left[
 \left( g \frac{\ctw^2-\stw^2}{\ctw}  J_Z^\mu + 2 \stw \ctw e J_{em}^\mu \right)
 Z_\mu +  \frac{\ctw^2 g}{\sqrt 2} ( J_W^\mu W^+_\mu   + {\rm h.c} ) \right]  \, .
  \label{Og1Z2}
\end{eqnarray}
Here we have substituted $\stw  g' J_Y +  g \ctw  J^{3} = (g/\ctw)(\ctw^2-\stw^2) \, J_Z + 2 \stw \ctw e J_{em}$.
The polynomials $Q_1$ and $Q_2$ are given in \eq{Q1}, and we have introduced
\bea
S_2 &=& 1 +  \frac{4 h}{3 v} +   \frac{h^2}{3 v^2}     \ ,   \nn \\
Q_3 &=& 1 +   \frac{5 h}{2v} + \frac{2 h^2}{v^2} +   \frac{h^3}{2 v^3}    \ .
\label{Q3}
\eea
We define the direction \eq{Og1Z2}  so that the coefficient 
$\eta_{g1Z} = \delta g_1^Z = \delta \kappa_Z$ in \eq{Leff2}.

We have defined as one of our directions the custodial-preserving structure $ {\widehat {\cal D}}_{h(VV)_c} $ in \eq{ChVV}.
Obviously, we should allow for the possibility of a custodial-breaking term in the Lagrangian. This can be done
with any combination of $hW_\mu W^\mu$ and $hZ_\mu Z^\mu$ different from the one in \eq{ChVV}.
We choose $hZ_\mu Z^\mu$ as the custodial-breaking term -that is why it appears in \eq{Og1Z2}.
In our approach this is a simple way to describe custodial-symmetry breaking effects.

As we said, to obtain the operator ${\cal O}_{g1Z}$ in the coupling basis we have to combine a piece proportional to ${\cal O}_{h(VV)_c}$
in \eq{hVVr6} with ${\cal O}_{B}-{\cal O}_{W}$. The exact combination is
\be
{\cal O}_{g1Z}={\cal O}_{B}-{\cal O}_{W} - g^2 \left( {\cal O}_{r} - \frac{1}{2} {\cal O}_{6} \right) \ .
\ee
In the unitary gauge, ${\cal O}_{g1Z}$ gives $ {\cal D}_{g1Z}$  provided we
make some redefinitions in the SM Lagrangian.

\subsubsection{$\mathbf{ a=\kappa\, \gamma}$}

Now it is the turn of $\kappa_\gamma$, related to ${\cal O}_{HB}$,
\be
{\cal O}_{HB} \rightarrow i g H^2 W^-_\mu W^+_\nu B^{\mu \nu} + \frac{2}{\ctw} H \, (\partial_\nu  H ) \, Z_\mu B^{\mu \nu} \ .
\ee
Indeed, the first term in the rhs contains the coupling $\widehat {\cal D}_{\kappa\gamma}$ (for $H=v$).
After partial integration we writte the second term as
\be
\frac{1}{2\ctw} (H^2 - v^2)    Z_{\mu \nu} B^{\mu \nu} - \frac{1}{\ctw} (H^2 - v^2)  Z_\mu \partial_\nu B^{\mu \nu} \ .
\label{Okg1}
\ee
In this last equation, the first term contains the vertex $h\gamma Z$ and it has to be subtracted using a  combination
of ${\cal O}_{BB}$ and ${\cal O}_{WW}$,
see \eq{hgghgZ}.
In the second term in \eq{Okg1} we use
the EoM \eq{eomV}. After some algebra one obtains
\begin{eqnarray}
{\cal D}_{\kappa\gamma} & = & i e \, R_2 \, W^+_\mu W^-_\nu  \left[    A^{\mu \nu} - \ttw    Z^{\mu \nu} \right] 
+   \left( \frac{h}{v} P_1 \right) \left[ 
\cW^+_{\mu \nu} \cW^{- \, \mu \nu} +	\frac{\ctw^2 - \stw^2}{2 \ctw^2} \cZ_{\mu \nu} \cZ^{\mu \nu} 
\right. \nn \\ 
&+& \left. 
\ttw \left(\cA_{\mu \nu} \cZ^{\mu \nu} -     A_{\mu \nu}     Z^{\mu \nu} \right) 
+ \ttw^2      Z_{\mu \nu}     Z^{\mu \nu} \right]
-   \frac{g'^2}{2\, \ctw^2}   \left( h Q_3 \right) v \, Z^\mu Z_\mu
\nn \\
&+& 
  2 \, \frac{g' \stw}{\ctw^2}  \left( \frac{h}{v} P_1\right) ( \ctw^2 J^\mu_{em} - J^\mu_Z  ) Z_\mu \ ,
  \label{Dkappag}
\end{eqnarray}
where $P_1$ and $Q_3$ are given in \eq{P1} and  \eq{Q3}, respectively, and
\be
R_2 = 1 +  \frac{2 h}{v} +  \frac{h^2}{v^2} 
\label{R2}
\ee
We have defined the direction \eq{Dkappag} in a way that, when summed in  \eq{Leff2},
is to be multiplied by $\eta_{\kappa\gamma} = \delta \kappa_\gamma = - \ttw^{-2} \delta \kappa_Z$.

We stress again that we choose the vertex $hZ_\mu Z^\mu$, appearing in the direction
\eq{Dkappag}, as a custodial-breaking term. 
In addition, notice the appearance of terms $hZ_{\mu\nu} Z^{\mu\nu}$ and $h W_{\mu\nu} W^{\mu\nu}$,
which contribute to $h \to V V^{(*)}$, also in a custodial-breaking combination.

The $d=6$ operator in the coupling basis is
\be
{\cal O}_{\kappa\gamma}= {\cal O}_{HB} - \frac{1}{8} \left( {\cal O}_{WW} - {\cal O}_{BB} \right) \ .
\ee
It gives the desired direction ${\cal D}_{\kappa\gamma}$ provided that at the same time some redefinitions of SM parameters are made.

\subsubsection{$\mathbf{ a=Z\, f\, f}$}
\label{sectionZff}

Let us now turn our attention to operators with fermions. 
We start with the seven operators in \eq{second6dimB},
which can be related in a quite direct way to the seven couplings in \eq{dirZff1} and \eq{dirZff2}.

The directions $a=ZfR$, $f=u,d,e$  in  \eq{dirZff1} are easily to obtain because 
${\cal O}_{ZfR} = {\cal O}^f_{R} $, so that
\begin{eqnarray}
{\cal D}_{Z dR} & = &  R_2  \ Z_\mu \ \bar d_R \gamma^\mu d_R  \ ,  \nonumber \\
{\cal D}_{Z uR} & = & R_2 \ Z_\mu \ \bar u_R \gamma^\mu u_R \ ,  \nonumber  \\
{\cal D}_{Z eR} & = &  R_2  \ Z_\mu \ \bar e_R \gamma^\mu e_R \ .
\end{eqnarray}
Suitable combinations of ${\cal O}^l_{L}, {\cal O}_{L}^{(3)\, l}, {\cal O}^q_{L}, {\cal O}_{L}^{(3)\, q}$
give the $a=ZfL$, $f=u,d,\nu,e$, directions in \eq{dirZff2},
Indeed,
\be
{\cal O}_{Z uL} = {\cal O}^{(3) q}_{L} -  {\cal O}^{q}_{L}   \ , \ \
{\cal O}_{Z dL} = {\cal O}^{(3) q}_{L} +  {\cal O}^{q}_{L}  \ ,
\ee
and similarly for $e$ and $\nu$. We get
\begin{eqnarray} 
{\cal D}_{Z uL} & = &  R_2  \left[ Z_\mu \ \bar u_L \gamma^\mu u_L +  
\frac{\ctw}{\sqrt{2} }  \left( W^+_\mu \ \bar u_L \gamma^\mu d_L + {\rm h.c.} \right) \right]  \ , \nonumber  \\
{\cal D}_{Z dL} & = & R_2  \left[ Z_\mu \ \bar d_L \gamma^\mu d_L -
\frac{\ctw}{\sqrt{2} }  \left(  W^+_\mu \ \bar u_L \gamma^\mu d_L + {\rm h.c.} \right) \right]   \ ,
\nonumber  \\
{\cal D}_{Z eL} & = & R_2   \left[ Z_\mu \ \bar e_L \gamma^\mu e_L - 
\frac{\ctw}{\sqrt{2} }  \left(  W^+_\mu \ \bar \nu_L \gamma^\mu e_L + {\rm h.c.} \right) \right]  \ ,
\nonumber  \\
{\cal D}_{Z \nu L} & = & R_2   \left[ Z_\mu \ \bar \nu_L \gamma^\mu \nu_L +
\frac{\ctw}{\sqrt{2} } \left(  W^+_\mu \ \bar \nu_L \gamma^\mu e_L + {\rm h.c.} \right) \right] \ ,
\end{eqnarray}
where $R_2$ is given in \eq{R2}. 

\subsubsection{$\mathbf{ a=D\, V\, f}$ (Dipoles)}

We now turn our attention to the dipole operators in \eq{second6dimC}
and the dipole couplings in \eq{couplings Dipole}.
For the gluons, we simply have ${\cal O}_{DGq} = {\cal O}^q_{DG}$. Thus,
\be
{\cal D}_{DGq}  = \left(  \frac{P_1}{v} \right) \left[ \bar q_L \sigma^{\mu  \nu} T^A q_R \, \cG^A_{\mu \nu}+ {\rm h.c.} \right] \ , 
\label{aDGq}
\ee
with $q=u,d$. Concerning the electroweak dipoles, we can reorganise the gauge invariant operators in \eq{second6dimC}
so that we obtain dipoles in the physical gauge-boson 
basis. We have, for the up-quark type
\be
{\cal O}_{DZu} = {\cal O}^{u}_{DW} -  {\cal O}^{u}_{DB}   \ , \ \
{\cal O}_{DAu} =  {\cal O}^{u}_{DB} + \ttw^2   {\cal O}^{u}_{DW}  \ ,
\ee
and for down-type quark
\be
{\cal O}_{DZd} =  {\cal O}^{d}_{DW} +  {\cal O}^{d}_{DB}   \ , \ \
{\cal O}_{DAd} =  {\cal O}^{d}_{DB} - \ttw^2   {\cal O}^{d}_{DW}  \ ,
\ee
and similarly for the electron $e$. The corresponding directions ${\cal D}_{DVf}$ are
\bea
{\cal D}_{DVu} &=& \left(  \frac{P_1}{v}  \right) \left[ \bar u_L \sigma^{\mu \nu} u_R \cV_{\mu \nu} + 
\sqrt{2} \, c_V \left( \bar d_L \sigma^{\mu \nu} u_R \, \cW^-_{\mu \nu} + {\rm h.c.} \right)  \right]  \ ,  \nn \\
{\cal D}_{DVd} &=& \left(  \frac{P_1}{v}  \right) \left[ \bar d_L \sigma^{\mu \nu} d_R \cV_{\mu \nu} - 
\sqrt{2}  \left(  c_V \,  \bar d_L \sigma^{\mu \nu} u_R \, \cW^-_{\mu \nu} + {\rm h.c.} \right)  \right]
\ ,  \nn \\
{\cal D}_{DVe} &=& \left(  \frac{P_1}{v}  \right) \left[ \bar e_L \sigma^{\mu \nu} e_R \cV_{\mu \nu} -
\sqrt{2} \left(  c_V \, \bar e_R \sigma^{\mu \nu} \nu_L \, \cW^-_{\mu \nu} + {\rm h.c.}  \right)  \right]  \ .
\label{aDVf}
\eea
Here $\cV= \cZ,  \cA$, as defined in \eq{amunu}, and $c_Z= \ctw, c_A= \stw$.
One may easily extend the results to three families. 

\subsubsection{$\mathbf{ a=uR\, dR}$}

The operator in \eq{second6dimD} leads to
\be
{\cal O}_{uRdR} = {\cal O}_{R}^{ud} \rightarrow H^2 \,  \bar u_R \gamma^\mu d_R W^+_\mu 
\ee
and to
\be
{\cal D}_{uRdR} =  R_2 \, \bar u_R \gamma^\mu d_R W^+_\mu + {\rm h.c.}
\ee

\subsection{TGC-only Sector}
\label{tgcOnly}

\subsubsection{$\mathbf{ a=\lambda\, V}$}

The operator \eq{3W}
 induces a TGC coupling accompanied by quartic interactions.
The direction  ${\cal O}_{\lambda V}  = {\cal O}_{3W}$ leads to
\be
{\cal D}_{\lambda V} =  \frac{i}{m^2_W}  \left( g  \ctw \, \cZ^{\mu \nu} + e \, \cA^{\mu \nu} \right)  
\cW^{- \rho}_{\nu} \, \cW^+_{ \rho \mu}   \ .
\label{OlambdaV}
\ee
The normalisation is $\eta_{\lambda V}= \lambda_Z = \lambda_\gamma$, in the
parameterization in \eq{V3}. 

\section{The CP-odd Directions}
\label{odd}

The directions we associate to   \eq{Oodd1} are $a= hg \widetilde g, h\gamma \widetilde \gamma, h\gamma \widetilde Z$.
To get them we use the decomposition \eq{decomposition}. The terms with $v^2$ can be obviated, because they can be
reabsorbed in the cases of  ${\cal O}_{B \widetilde{B}}$ and ${\cal O}_{W \widetilde{W}}$, and in the case of $ {\cal O}_{G \widetilde{G}}$
because it is a contribution to the $\theta$-QCD parameter, which we assume is rotated away with, for instance, a Peccei-Quinn  mechanism.
It follows that the operators  \eq{Oodd1}  are of the "Higgs-only" CP-odd type. 

The direction corresponding to the CP-odd coupling of Higgs to two gluons 
is simply given by ${\cal O}_{hg \widetilde g}  = {\cal O}_{G\widetilde G}$, and from here we obtain
\be
{\cal D}_{ hg \widetilde g } = \left( \frac{h}{v} P_1 \right) \cG_{\mu\nu}^A \widetilde \cG^{A\mu\nu} \ .
\ee
 The CP-odd coupling of Higgs to $\gamma \gamma, \gamma Z$ can be obtained with the simple linear combinations 
 ${\cal O}_{h\gamma\widetilde \gamma} = {\cal O}_{B\widetilde B} +\ttw^2  {\cal O}_{W\widetilde W}$,
and 
$ {\cal O}_{h\gamma\widetilde \gamma}  =
 {\cal O}_{h\gamma \widetilde Z}   =    {\cal O}_{B\widetilde B} -  {\cal O}_{W\widetilde W} $.
We get
 \bea
{\cal D}_{h\gamma \widetilde \gamma} &=& \left( \frac{h}{v} P_1 \right)  \left[ \cA_{\mu \nu}  \widetilde \cA^{\mu \nu} + \cZ_{\mu \nu} 
 \widetilde \cZ^{\mu \nu} +  2\, \cW^+_{\mu \nu}  \widetilde \cW^{- \, \mu \nu}   \right] \nn \\
 {\cal D}_{h\gamma   \widetilde  Z} &=&  \left( \frac{h}{v} P_1 \right)  \left[ \cA_{\mu \nu}  \widetilde \cZ^{\mu \nu} + 
 \frac{\ctw^2 - \stw^2}{2 \ctw \stw} \cZ_{\mu \nu}  \widetilde \cZ^{\mu \nu} + 
\frac{\ctw}{\stw} \, \cW^+_{\mu \nu}  \widetilde  \cW^{- \, \mu \nu}   \right] \ .
\eea

The two operators in \eq{Oodd2} contribute to the CP-odd TGC, see \eq{V3odd}.
Consider first  $a= \widetilde{\kappa \gamma}$.
When working out ${\cal O}_{H\widetilde B}$ in \eq{Oodd2} one finds that the
vertex corresponding to $h\gamma \widetilde Z$ has to be subtracted, like
in the CP-even case in \eq{Okg1}. However, a difference with the CP-even direction
is that there are no more subtractions to be done because $\partial^\nu \widetilde B_{\mu \nu}=0$.
As a result,
\begin{eqnarray}
{\cal D}_{ \widetilde{\kappa \gamma} } & = & i e\,  R_2 \, W^+_\mu W^-_\nu  \left[    \widetilde  A^{\mu \nu} - 
\ttw   \widetilde   Z^{\mu \nu} \right]  +  \left( \frac{h}{v} P_1 \right) \left[ 
\cW^+_{\mu \nu}  \widetilde \cW^{- \, \mu \nu} +	\frac{\ctw^2 - \stw^2}{2 \ctw^2} \cZ_{\mu \nu}  \widetilde \cZ^{\mu \nu} 
\right. \nn \\ 
&+& \left. 
\ttw \left(\cA_{\mu \nu}  \widetilde \cZ^{\mu \nu} -     A_{\mu \nu}    \widetilde   Z^{\mu \nu} \right) 
+ \ttw^2      Z_{\mu \nu}     \widetilde  Z^{\mu \nu} \right] \ ,
\end{eqnarray}
where the normalisation is such that $\eta_{\widetilde{\kappa \gamma} }= \delta \widetilde{\kappa}_\gamma$.
The relation between  bases in this case is
${\cal O}_{ \widetilde{\kappa \gamma} } = {\cal O}_{H\widetilde B} - ({\cal O}_{W\widetilde W}  - {\cal O}_{B\widetilde B} )/8$

The contribution to $ a = \widetilde {\lambda V}$
comes directly from ${\cal O}_{3\widetilde W}$ in \eq{Oodd2}, so that
${\cal O}_{ \widetilde{\lambda V} } = {\cal O}_{3\widetilde W}$ and
\be
{\cal D}_{\widetilde {\lambda V}} =  \frac{i}{m^2_W}  \left(g  \ctw \, \cZ^{\mu \nu} + e \, \cA^{\mu \nu} \right   )  
\cW^{- \rho}_{\nu} \, \widetilde \cW^+_{ \rho \mu}   \ .
\label{OlambdaVtilde}
\ee

\section{The coupling basis}
 \label{observablebasis}
 
In Sections \ref{even} and \ref{odd}, for the chosen set of couplings $\{ a \}$,
 we have determined the directions $\{ {\cal D}_{a} \}$ and 
the linear combinations of the ${\cal O}_{i}$'s we have to perform 
to get the coupling basis. Here we summarise all of them with a list of results in the form
 \be
 {\cal O}_{a}   \ = \ \sum_i  \kappa^i_a {\cal O}_{i} \ \rhd \ \kappa_a \, {\cal D}_{a}   \ , 
 \ee
 where  $\kappa^i_a$ shows the required particular combination for each $a$, and $\kappa_a$ gives
 the normalisation. The symbol $\rhd$ means that we take the expressions for ${\cal O}$ in the unitary gauge
 and that, when needed, we redefine some SM bare parameters.
 
We  first list the operators that do no involve fermions (even if fermions are present in some directions),
 \bea
 {\cal O}_{h3}    \   =& {\cal O}_{6} &\rhd  \  \ \lambda \, v^2 \, {\cal D}_{h3} \nn   \ ,  \\
 {\cal O}_{h(VV)_c}     \ =&     {\cal O}_{r} - \frac{1}{2} {\cal O}_{6} &\rhd \   \ \frac{g^2}{4}  \, v^2 \,  {\cal D}_{h(VV)_c}    \nn   \ ,  \\
  {\cal O}_{hgg}    \   =&   {\cal O}_{GG} &\rhd \  \  g_s^2  \, v^2 \,  {\cal D}_{hgg}   \nn   \ ,  \\
 {\cal O}_{h\gamma\gamma}      \  =&    {\cal O}_{BB} +\ttw^2  {\cal O}_{WW} &\rhd \  \  g'^2  \, v^2 \,  {\cal D}_{h\gamma\gamma}    \ ,  \nn  \\
 {\cal O}_{h\gamma Z}     \  =&    {\cal O}_{WW} -  {\cal O}_{BB} &\rhd \  \ 2  g g'   \, v^2 \,  {\cal D}_{h\gamma Z}   \ ,  \nn  \\
 {\cal O}_{g1Z}     \  =&    {\cal O}_{B} -  {\cal O}_{W} -g^2 \left( {\cal O}_{r} - \frac{1}{2} {\cal O}_{6} \right) &\rhd \   \ 
 \frac{g^2}{4 \, \ctw^2}   \, v^2 \,  {\cal D}_{g1Z}    \ ,  \nn  \\
 {\cal O}_{\kappa \gamma}     \   =&    {\cal O}_{HB} - \frac{1}{8} \left(  {\cal O}_{WW}  - {\cal O}_{BB} \right) &\rhd \ 
 - \, \frac{g^2}{4}   \, v^2 \,  {\cal D}_{\kappa \gamma}   \ ,  \nn \\
  {\cal O}_{\lambda V}      \  =&    {\cal O}_{3W}  &\rhd \  \   \frac{g^2}{4}   \, v^2 \,  {\cal D}_{\lambda V}   \ ,
  \label{table1}
 \eea
 and operators with fermions
  \bea
 {\cal O}_{hff}        \    =&  {\cal O}_{y_f} &\rhd \    \ \frac{y_f}{\sqrt{2}}  \, v^2 \,  {\cal D}_{hff} \nn   \ ,  \\
 {\cal O}_{ZfR}     \       =&     {\cal O}_{R}^f &\rhd \  - \, \frac{g}{2\ctw}  \, v^2 \,  {\cal D}_{ZfR}   \ ,  \nn  \\
 {\cal O}_{ZfL}       \     =&    {\cal O}_{L}^{(3) F} -  (2 T_3^f)\, {\cal O}_{L}^{F} &\rhd \  \ (2 T_3^f)\, \frac{g}{\ctw}  \, v^2 \,  {\cal D}_{ZfL}   \ ,  \nn  \\
 {\cal O}_{DGq}      \      =&  {\cal O}_{DG}^q &\rhd \   \  y_q g_s  \, v^2 \,  {\cal D}_{DGq}    \ , \nn  \\
 {\cal O}_{DZf}       \        =& {\cal O}_{DW}^f   - (2T_3^f )\, {\cal O}_{DB}^f &\rhd \  \ (2T_3^f )\,  y_f \frac{g}{\ctw}  \, v^2 \,  {\cal D}_{DZf}   \ , \nn  \\
 {\cal O}_{DAf}       \     =&  {\cal O}_{DB}^f   + (2T_3^f)\,  \ttw^2 {\cal O}_{DW}^f &\rhd \   \ y_f \frac{g'}{\ctw}  \, v^2 \,  {\cal D}_{DAf}   \ ,  \nn  \\
 {\cal O}_{uRdR}      \      =&  {\cal O}^{ud}_R &\rhd \    \  \frac{y_u y_d}{\sqrt{2}} g^2 \, v^2 \,  {\cal D}_{uRdR}   \ .
   \label{table2}
\eea
Here, $f=u,d,e$ (and also $\nu$ in $ZfL$),  $F=q,l$, and $q=u,d$. The weak isospin of $f$
is denoted by $T_3^f $.
In the dipole operators, the operators ${\cal O}_i$ are understood to be summed with its hermitian conjugate.

The CP-odd  list  is 
\bea
{\cal O}_{hg\widetilde g}    \   =&   {\cal O}_{G\widetilde G} &\rhd \   \ g_s^2  \, v^2 \,  {\cal D}_{hg\widetilde g}     \ , \nn  \\
{\cal O}_{h\gamma\widetilde \gamma}     \   =&    {\cal O}_{B\widetilde B} +\ttw^2  {\cal O}_{W\widetilde W} &
\rhd \   \ g'^2  \, v^2 \,  {\cal D}_{h\gamma\widetilde \gamma}    \ ,  \nn  \\
 {\cal O}_{h\gamma \widetilde Z}     \  =&    {\cal O}_{W \widetilde W} -  {\cal O}_{B\widetilde B} &\rhd \  \ 2 g g'   \, v^2 \,  {\cal D}_{h\widetilde \gamma Z}   \ ,  \nn  \\
 {\cal O}_{\widetilde {\kappa \gamma}}     \   =&    {\cal O}_{H\widetilde B} - \
 \frac{1}{8} \left(  {\cal O}_{W\widetilde W}  - {\cal O}_{B\widetilde B} \right) &\rhd \ 
 - \, \frac{g^2}{4}   \, v^2 \,  {\cal D}_{\widetilde {\kappa \gamma}}   \ , \nn \\
  {\cal O}_{\widetilde {\lambda V}}    \    =&    {\cal O}_{3\widetilde W}  &\rhd \   \  \frac{g^2}{4}   \, v^2 \,  {\cal D}_{\widetilde {\lambda V}}   \ .
  \label{tableCPodd}
 \eea

We stress that the expressions for the ${\cal D}_{a}$ are unique, up to partial integrations and redefinitions. 
The particular linear combinations that lead
from $\{ {\cal O}_{i} \}$ to $\{ {\cal O}_{a} \}$ depend of course on the particular starting basis we have chosen. 
Just as an example, if in our starting basis we had the operator ${\cal O}_H = \frac{1}{2}(\partial^\mu |\H|^2)^2$
instead of ${\cal O}_r$, we could also get the $h(VV)_c$ direction in \ref{hVVsubS}, but using the combination
 \be
 {\cal O}_{h(VV)_c}  \ = \ \ {\cal O}_{H} - \sum_f \frac{1}{2} {\cal O}_{y_f} - \frac{3}{2} {\cal O}_{6} \ \rhd  \ - \, \frac{g^2}{4}  \, v^2 \,  {\cal D}_{h(VV)_c}
 \ee
We see that the combinations required to get $a=h(VV)_c$ in the case of ${\cal O}_r$ are simpler than in the case of ${\cal O}_H$,
and this is why we opted for the first. This is an example of choosing
 the starting basis as ``close" as possible to $\{ {\cal O}_{a} \}$.

A final comment is that we have systematically eliminated the presence of terms with derivatives acting on the Higgs field.
In our case, this can be done using partial integration and applying EoMs.

\section{Discussion and Conclusions}
\label{conclusion}

The correlations induced
by gauge-invariant operators in effective Lagrangians have
interesting phenomenological consequences. In the case of
observing a deviation from a SM prediction -sign of BSM-, 
such correlations  imply that deviations on some other couplings
should be found.
By the same token, constraints on deviations from the SM in some couplings imply constraints
on other couplings. The correlations might be a very useful guide to BSM physics.
However, the analysis is complicated by the fact that, in general, each operator contributes to different couplings 
and, in turn, each coupling receives
constraints from several operators.

In this paper we have shown how to write the effective Lagrangian in a basis where each independent 
gauge-invariant operator ${\cal O}_a$
is linked to a particular coupling $a$, described by a term $ \widehat{ {\cal D}}_a$ multiplied by
$\eta_a$, the potential deviation from the SM prediction. In a sense, our work has been to "complete" $ \widehat{ {\cal D}}_a$
with other structures to have finally a gauge-invariant operator  ${\cal O}_a$. 
Each term  $\widehat{{\cal D}}_a$ appears once, and only once, in the effective Lagrangian.
With this requirements, the expressions for ${\cal O}_a$ and ${\cal D}_a$ are unique, up to partial integrations, redefinitions etc.

The number of couplings $\{ a \}$ in the sum \eq{Leff2}  equals the number of operators
of a basis in operator space. We restrict our study to physics relevant for EWPT, TGC and Higgs physics. Then
this number is equal to 27 (in the CP-even sector, for one family) and 5 (in the CP-odd sector). This makes a total of 32.
We have chosen the couplings that are better measured or eventually will be. In the case that
in the future the experimental situation changes, we could easily choose another set of couplings
and work out a form of the effective Lagrangian suited to this novel group.

{{\renewcommand{\arraystretch}{1.35} 
\begin{table}[h] 
\begin{center}
\begin{tabular}{|c|c|c|}
\hline 
$a$   &
$\widehat {\cal D}_a$    &
$\delta {\cal D}_a$
\\ 
\hline
$hff \ (3) $ & $ h \bar f f  $ & $ h^2 \bar f f \, . $  \\
$ h3 $ &  $ h^3  $  &  $ h^4 \, .$    \\
$h(VV)_c  $ & $  vh  \left(  W^{+\mu} W^-_\mu + \frac{1}{2 \ctw^2}  Z^\mu Z_\mu \right)  $ 
&  $ h   \widehat {\cal D}_{h(VV)_c} \, , \ h^2 \bar f f \, , \ h^4 \, .$    \\
$h\gamma\gamma  $ &  $ \frac{1}{v} \, h  A^{\mu \nu} A_{\mu \nu}  $    
& $  h   \widehat {\cal D}_{h\gamma\gamma} \, , \  h W^2 \partial V_{\gamma,Z} \, ,  $      \\
 &   & $ hW V_{\gamma,Z} \partial W \, , \  h^{1,2}   (\partial V_{Z,W})^2   \, . $       \\
$h\gamma Z  $ &  $ \frac{1}{v}   \, h A^{\mu \nu} Z_{\mu \nu}    $ 
&  $ h   \widehat {\cal D}_{h\gamma Z} \, , \  h W^2 \partial V_{\gamma,Z} \, , $ \\
 &   & $ hW V_{\gamma,Z} \partial W \, , \  h^{1,2}  (\partial V_{Z,W})^2  \, . $    \\
 \hdashline
 $g1Z  $ &  $ i  g \ctw  [ Z^\mu ( W^{+\nu} W^-_{\mu \nu } - {\rm h.c})    $ 
& $  ZW^2 V_{\gamma,Z} \, , \ W^4\, , \  h^2 (VV)_c \, , \ h^{1,2} (VV)_{nc} \, ,$  \\
    & + $ Z^{\mu \nu} W^+_\mu W^-_\nu  ]    $ 
&  $  h^2\bar f f\, , \ h^4 \, , h Z J_{R,L} \, , \  h W J_L \, .$  \\
$\kappa\gamma$ & $i e    W^+_\mu  W^-_\nu ( A^{\mu\nu} - \ttw Z^{\mu\nu} )$  
& $ h   \widehat {\cal D}_{\kappa\gamma} \, , \ h W^2 \partial V_{\gamma, Z} \, , \  h V_{\gamma, Z} W \partial W  \, ,
  $ \\
    &    
&  $ h^{1,2} (\partial V_{Z,W})^2 \, , \ h^{1,2} (VV)_{nc}\, , \ h Z J_{R,L}  \, . $ \\
 \hdashline
 $ZfR$ \ (3) & $Z_\mu \bar f_R \gamma^\mu f_R $ &
 $ h \widehat {\cal D}_{ZfR}  \, .  $    \\
 $ZfL$ \ (4) & $Z_\mu \bar f_L \gamma^\mu f_L $ &
 $h \widehat {\cal D}_{ZfL}$ \,  ,  \ $h W J_L  \, .  $   \\
\hline
\end{tabular}
\caption{\it For each direction $a$ and  $\widehat {\cal D}_{a} $ we show the couplings in $\delta {\cal D}_{a} $ 
involving three or four particles.
We display them schematically, for example $\partial V$ stands for the $V_{\mu \nu}$ field strength, 
$V_{\gamma, Z}$ for $A_\mu$ or $Z_\mu$, $Z J_R$ for $Z_\mu \bar f_R \gamma^\mu f_R$, 
$h^{1,2}$ for $h$ and $h^2$, etc. We write the non-custodial coupling $hZ^\mu Z_\mu$ as $h(VV)_{nc}$.
The exact form of the couplings as well as the factor in front are given
in the formulas in the text. The 16 directions in the Table form a group with the following property.
Any of them has at least one
common term with at least another of direction in the group. In the first column, the number
is parenthesis corresponds to the number of directions, when greater than one (for one fermion family). 
In $a=hff,ZfR$, $f=u,d,e$; in $a=ZfL$, $f=u,d,e,\nu$
The dashed lines separate 
directions measured in $h$-physics, TGC and $Z$-physics.}
\label{table:one}
\end{center}
\end{table}
}

In Tables 1, 2 and 3 we list all the couplings in the set $\{ a \}$, together with the defining operator $ \widehat{ {\cal D}}_a$.
The complete direction is $ {\cal D}_a= \widehat{ {\cal D}}_a + \delta {\cal D}_a$. In the Tables we show schematically
which couplings one has for each $a$, up to four-particle vertices. Given two different couplings $a$ and $b$ we
may ask whether there are some common terms in $ \delta {\cal D}_a$ and $\delta {\cal D}_b$. This has physical
interest, because a common term in two directions means the existence of a correlation. More generally, we may
separate our set of couplings $\{ a \}$ in groups or classes, such that inside a group  each direction
has at least a common term with at least another direction of the group. In other words, a direction in  a group
has no common term with any other direction of the other groups. This splitting leads to a group of 16 (CP-even) directions
shown in Table 1. The 11 remaining CP-even directions are split in 7 classes, indicated in Table 2.
Finally, in Table 3 we show the 5 CP-odd directions, separated in 4 classes.

{{\renewcommand{\arraystretch}{1.35} 
\begin{table}[h] 
\begin{center}
\begin{tabular}{|c|c|c|}
\hline 
$a$   &
$\widehat {\cal D}_a$    &
$\delta {\cal D}_a$
\\ 
\hline
 $h gg$ & $\frac{1}{v}   \, h G^{A \mu \nu} G^A_{\mu \nu} $  
& $h   \widehat {\cal D}_{gg} \,  ,\ h G^2 \partial G \, .$ \\
 \hline
$\lambda V$ &
$\frac{i}{m^2_W}  \left( g  \ctw \, Z^{\mu \nu} + e \, A^{\mu \nu} \right)  
W^{- \rho}_{\nu} \, W^+_{ \rho \mu}$ & $W^2 (\partial W)^2$\, , \ $V_{\gamma ,Z}  W \partial V_{\gamma , Z}  \partial W  \, .$ \\
  \hline
 $DGu$ & $\frac{1}{v}   \, \bar u \sigma^{\mu \nu} T^A u \, G^A_{\mu \nu} $  
& $h   \widehat {\cal D}_{DGu}  \, .$  \\
\hline
 $DGd$ & $\frac{1}{v}   \, \bar d \sigma^{\mu \nu} T^A d \, G^A_{\mu \nu} $  
& $h   \widehat {\cal D}_{DGd}  \, .$\\
\hline
 $DVq$ \ (4) & $\frac{1}{v}   \,  \bar q \sigma^{\mu \nu} q \, V_{\mu \nu} $  
& $h   \widehat {\cal D}_{DVq} \, , \ (\bar q \sigma^{\mu \nu} q)\,  W^+_\mu  W^-_\nu \, , $ \\
 &  & 
 $h^{0,1} (\bar u  \sigma^{\mu \nu} d) \, W_{\mu \nu} \, , \  ( \bar u  \sigma_{\mu \nu} d) \, W^\mu V^\nu_{\gamma, Z} \, .$ \\
\hline
 $DVe$ \ (2) & $\frac{1}{v}   \, (\bar e \sigma^{\mu \nu} e)\, V_{\mu \nu} $  
& $h   \widehat {\cal D}_{DVe}  \, , \ (\bar e \sigma^{\mu \nu} e)\,  W^+_\mu  W^-_\nu \, , $  \\
&  & $  h^{0,1} (\bar e  \sigma^{\mu \nu} \nu)\, W_{\mu \nu} \, , \ (\bar e  \sigma_{\mu \nu} \nu) \, W^\mu V^\nu_{\gamma, Z} \, .$  \\
\hline
 $uRdR$ &  $ \bar u_R \gamma^\mu d_R W^+_\mu  $  & $h   \widehat {\cal D}_{uRdR}  \, .$ \\
\hline
\end{tabular}
\caption{\it  Same than Table 1, for the rest of CP-even directions. Here the split is in 7 groups. 
In $a=DVq,DVe$ we have $q=u,d$, $V=A,Z$.}
\label{table:one}
\end{center}
\end{table}
}

{{\renewcommand{\arraystretch}{1.35} 
\begin{table}[h] 
\begin{center}
\begin{tabular}{|c|c|c|}
\hline 
$a$   &
$\widehat {\cal D}_a$    &
$\delta {\cal D}_a$
\\ 
\hline
 $h g\widetilde g$ & $\frac{1}{v}   \, h G^{A \mu \nu} \widetilde G^A_{\mu \nu} $  
& $h   \widehat {\cal D}_{g\widetilde g} \,  ,\ h G^2 \widetilde{ \partial G} \, .$ \\
 \hline
$h\gamma \widetilde \gamma  $ &  $ \frac{1}{v} \, h  A^{\mu \nu} \widetilde A_{\mu \nu}  $    
& $  h   \widehat {\cal D}_{h\gamma\widetilde \gamma} \, , \  h W^2 \widetilde{\partial  V}_{\gamma,Z} \, ,  $      \\
 &   & $ hW V_{\gamma,Z} \widetilde{ \partial W} \, , \  h^{1,2}   \partial V_{Z,W}  \widetilde{ \partial V}_{Z,W} \, . $       \\
$h\gamma \widetilde Z  $ &  $ \frac{1}{v}   \, h A^{\mu \nu} \widetilde Z_{\mu \nu}    $ 
&  $ h   \widehat {\cal D}_{h\gamma \widetilde Z} \, , \  h W^2 \widetilde{\partial  V}_{\gamma,Z} \, , $ \\
 &   & $ hW V_{\gamma,Z} \widetilde{ \partial W} \, , \  h^{1,2}  \partial  V_{Z,W} \widetilde{ \partial V}_{Z,W}  \, . $    \\
  \hline
$ \widetilde{ \kappa\gamma}$ & $i e    W^+_\mu  W^-_\nu ( \widetilde A^{\mu\nu} - \ttw \widetilde Z^{\mu\nu} )$  
& $ h   \widehat {\cal D}_{\widetilde{ \kappa\gamma}} \, , \ h W^2 \partial \widetilde{ V}_{\gamma, Z}   \, ,
  $ \\
    &    
&  $ h V_{\gamma, Z} W \widetilde{ \partial W} \, , \ h^{1,2} \partial V_{Z,W} \widetilde{ \partial V}_{Z,W} \, .  $ \\
   \hline
$\widetilde{\lambda V}$ &
$\frac{i}{m^2_W}  \left( g  \ctw \, Z^{\mu \nu} + e \, A^{\mu \nu} \right)  
W^{- \rho}_{\nu} \, \widetilde W^+_{ \rho \mu}$ & $W^2 \partial W \widetilde{ \partial W}$\, , \ 
$V_{\gamma ,Z}  W \partial V_{\gamma , Z}  \widetilde{ \partial W}  \, .$ \\
  \hline
\end{tabular}
\caption{\it  Same than Table 1, for the CP-odd directions. }
\label{table:one}
\end{center}
\end{table}
}

A first good property of our approach has to do with what we have just discussed: 
In the search for deviations from the SM predictions, our effective Lagrangian \eq{Leff2} 
may facilitate the analyses looking for BSM signals. There are more advantages of our approach that we now comment.

In general,  effective Lagrangians induce corrections to the fine structure constant $\alpha$, and to masses of
particles in the SM.
When we use, for example $\alpha$ and $m_Z$ in the set of input parameters, 
those corrections propagate to the predictions of the effective Lagrangian.
(These effects were called indirect corrections in \cite{Rujula:1991se}.)
They appear because the predictions of the SM have to be expressed as functions of $\alpha$, $m_Z$, etc.
In this regard, a positive aspect of our approach is that all the terms in  our ${\cal L}_6$ are vertices containing three or more
particles. This means that there are no corrections to tree-level masses. Also, there is no correction to 
$\alpha$. The form of our effective Lagrangian makes the use of $\alpha$, $m_Z$, $m_W$, $m_h$ and $m_f$
as input parameters very convenient. To use $m_W$ instead of $G_F$ is simpler, because the use of the latter
involves a four-fermion operator. Besides, the $m_W$ precision has reached an accuracy that allows to use
its measurement as input. 
When we use $\alpha$ and particles masses as input, we do not have indirect corrections, and thus calculations using the form of our
effective Lagrangian are simplified.

We should also mention the problem of blind directions \cite{Rujula:1991se}. 
It may happen that in a certain basis there is a linear combination of operators that experiments cannot
bound, i.e., a blind direction.  By definition, in our approach there are no blind directions.

Even if is not the purpose of this paper to look for concrete applications of our approach,
it may be useful to sketch ideas about some possible applications. As we said, 
in the case of measuring a deviation from the SM prediction in a certain coupling $a$, one
can immediately see in $\delta {\cal D}_a$  which vertices should show a positive signal, and with what
strength. Of course we knew there are correlations, but the expressions we get in our framework for the directions show them
in a transparent way. 

Another aspect, complementary of what we have just explained, is that our results
show when there is no correlation. Let us explain it with an example. The LHC is expected to improve the LEP-2 measurements
on TGC. Suppose that  there is no sign of new physics, and thus LHC improves
the constraints on TGC. We may ask the question:
 Is this going to tell us something about  $h \to \gamma\gamma$
or $h \to \gamma Z$ decays ? We cite this example because in a general  basis,
for example the one in \cite{Grzadkowski:2010es},
one has operators contributing both to TGC and these Higgs decays, so that it is difficult to see
whether there are correlations or not.
Thanks to our formalism, we can easily see that the TGC directions are completely independent of the $h\gamma\gamma$
and the $h  \gamma Z$ directions, so that the answer to the preceding question is a clear no.

Let us  briefly mention other possible applications. One of the uses of effective Lagrangians is the power
to anticipate a constraint on a process using current limits on couplings.
Which are the relevant couplings and which one (or ones) is the dominant one can be readily seen 
in the expressions for our directions. A simple example will help to clarify this point.
In the effective Lagrangian there is a contact term $h V_\mu \bar f \gamma^\mu f$, with $V=Z,W$ which
contributes to the decays $h \to V ff$ measured at LHC. Our results show that such a term appears in the directions
$a=ZfR, ZfL, g1Z, \kappa \gamma$. Then the anticipated constraint on $hVff$ comes from these terms.
In addition, since the limits on $\eta_{g1Z}, \eta_{\kappa \gamma}$ are order percent while
on $ZfR, ZfL$ are order permille, it is the former that sets the magnitude of the constraint to be expected
for $hVff$. This was already pointed out in \cite{Pomarol:2013zra}.
Another example is the custodial breaking terms $h Z^\mu Z_\mu$ appearing only in the
directions $g1Z, \kappa \gamma$. The limits on this two TGC directions put a limit to
the expected breaking of custodial symmetry in the $h V^{\mu} V_\mu$, $V=Z,W$, amplitudes.
There are also custodial breaking effects in the couplings $h V^{\mu\nu} V_{\mu\nu}$,
which appear in the directions $a=h\gamma Z, \kappa \gamma$.
An interesting final example refers to the quartic gauge couplings, which are starting being measured at LHC.
 In our work, we show precisely how the modifications
to the quartic gauge couplings are linked to the TGC parameters $g_1^Z, \kappa_\gamma$ and $\lambda_Z=\lambda_\gamma$.

Very recently, the Lagrangian \eq{LDa}  was deduced in \cite{Gupta:2014rxa}. The authors build
it in a bottom-up approach: they make use of symmetries and analyze field structures 
 to find all the independent terms in the effective Lagrangian, and generate as we have done
all directions corresponding to couplings. In  \cite{Gupta:2014rxa}, emphasis is done
on the fact that  \eq{Leff2} helps in the search of BSM physics, as we do in our paper.
There, the directions are called BSM Primary Effects.
The final results in \cite{Gupta:2014rxa} and the ones found here agree,
as expected.  The two methods are very different since we use a 
top-down approach. Indeed, we start with the full relevant set of $d=6$ operators and manipulate them to get the
directions. In a sense, our method is more systematic. As we said, in the future we could need to
 change the group of well-measured couplings. In this eventuality, we think our top-down approach is 
more suitable for such a change.

We end with a couple of remarks. First, our analytical results are not meant to be a substitute of a dedicated numerical
investigation, necessary to be able to state limits with a certain CL, etc. 
The Lagrangian we have reached helps us in the understanding of some general properties, but of 
course can be perfectly used for numerical analyses.
Second, our findings assume the framework of
a linear effective Lagrangian with $d=6$ operators. In the case that the predicted
 correlations are not observed, this would mean that such a framework is not valid, 
 which would also be precious information on BSM. 
  
\section*{Acknowledgments}

I am thankful to Alex Pomarol for illuminating discussions, and to Joan Elias-Miro, Christophe Grojean and
Francesco Riva for a critical reading of the typescript.
 This work is
supported by the CICYT Research Project FPA2011-25948 and by the
{\it Generalitat de Catalunya},
2014 SGR 1450.

\section*{Appendix A}

Here we write the EW SM Lagrangian, in the unitary gauge, in order to fix some notation.
We use the following conventions. For the covariant derivative
\be
 D_\mu = \partial_\mu - i g \, T^a W^a_\mu - i g' \, Y B_\mu  \ ,
\label{}
\ee
where $T^a=\sigma^a/2$ for $SU(2)$ doublets. 
 The Higgs field ($Y_\Phi=1/2$), in  the unitary gauge, reads
\be
 \Phi = \frac{1}{\sqrt{2}} \left(  \begin{array}{c} 0 \\ H \end{array} \right) =
 \frac{1}{\sqrt{2}} \left(  \begin{array}{c} 0 \\ v + h \end{array} \right)   \ .
\label{PhiDef}
\ee 
Here $H/\sqrt{2}$ is the neutral part of the Higgs doublet, $h$ is the physical Higgs field.

We write the  Lagrangian in the unitary gauge  as
\be
 {\cal L}_{SM} = {\cal L}_{G} + {\cal L}_{f} + {\cal L}_{Gf} + {\cal L}_H  \ .
\label{LSM}
\ee
The pure gauge and gauge-Higgs part is
\bea
{\cal L}_{G} &=&  - \, \frac{1}{4} \cA^{\mu\nu} \, \cA_{\mu\nu} \, - \, \frac{1}{4} \cZ^{\mu\nu} \, \cZ_{\mu\nu}  \, - \, \frac{1}{2} \cW^{+\mu\nu} \, \cW^-_{\mu\nu}
 \nonumber \\
 &+&  \frac{1}{4} g^2  H^2  \, W^{+\mu} W^-_\mu + \frac{1}{8}  \frac{g^2}{\ctw^2} H^2  \, Z^{\mu} Z_\mu \ .
 \label{LG}
\eea
We have defined
\bea
\cA_{\mu\nu} &=&    A_{\mu\nu} + i  g \stw  \left(  W^-_\mu   W^+_\nu - W^+_\mu   W^-_\nu \right)  \ , \nonumber \\
\cZ_{\mu\nu} &=&    Z_{\mu\nu} + i g \ctw  \left(  W^-_\mu   W^+_\nu - W^+_\mu   W^-_\nu \right) \ , \nonumber \\
\cW^\pm_{\mu\nu} & = &  {   W}^\pm_{\mu\nu} \pm i g  \left(  W^\pm_\mu   W^3_\nu - W^3_\mu   W^\pm_\nu \right)  \ ,
\label{amunu}
\eea
with $   V_{\mu\nu} = \partial_\mu V_\nu - \partial_\nu V_\mu$ (what we denote by $   A_{\mu\nu} $ is usually  denoted by $F_{\mu\nu} $).
We write the Lagrangian in terms of the weak bosons $W^\pm$, $Z$ and the photon $A$, and for that purpose we introduce
\be
\stw = \sin \theta_w = \frac{g'}{\sqrt{g^2 + g'^2}}  \ , \ \  \ctw = \cos \theta_w = \frac{g}{\sqrt{g^2 + g'^2}}  \ , \ \  \ttw = \tan \theta_w  \ .
\label{}
\ee
For example, in the definition of $\cW$ in \eq{amunu}, we have to substitute $W^3_\mu = \ctw Z_\mu + \stw A_\mu$.

The fermion and Higgs-fermion terms in (\ref{LSM}) are given by
\be
{\cal L}_{f} = i \bar f \slashed \partial f  - \frac{y_f}{\sqrt{2}} \, H  \bar f f  \ ,
\label{}
\ee
where $y_f$ are the Yukawa couplings and where we understand a sum over fermions. In (\ref{LSM}) we also have
the gauge boson-fermion interactions
\be
{\cal L}_{Gf} = \frac{g}{\sqrt{2}} \ ( J_W^\mu W^+_\mu   + h.c.) + \frac{g}{\ctw} \,  J_{Z}^\mu Z_\mu  + e \,  J_{em}^\mu A_\mu \ .
\label{LGf}
\ee
We define the currents 
\be
J^{a \mu}  = \bar F_L \frac{\sigma^a}{2} \gamma^\mu F_L   \qquad ,  \qquad   J_{Y}^\mu =  \bar f Y_f \gamma^\mu f  \ ,
\label{currents1}
\ee
where $F_L$ are the fermion doublets, and again a sum over fermions is understood. The currents appearing in (\ref{LGf}) are
\bea
J_W^\mu &=& J^{1 \mu} + i J^{2 \mu}  \ , \nn \\
 J_{Z}^\mu  &=& \ctw^2 J^{ 3\mu}  - \stw^2 J_{Y}^\mu   \ , \nn \\
 J_{em}^\mu &=& J^{3 \mu} + J_{Y}^\mu    = \bar f Q_f \gamma^\mu  f 
\label{currents2}
\eea
 Finally, the pure $H$-part is
\be
{\cal L}_{H} = \frac{1}{2}  (\partial^\mu H) (\partial_\mu H) -V(H)
\label{}
\ee
where the potential
\be
V(H) = \frac{\lambda }{4} \left( H^2 - v^2 \right)^2 = - \frac{ \mu^2}{2} H^2 + \frac{\lambda }{4} H^4 + {\rm constant}
\label{VH}
\ee

We take as input parameters $\alpha$ and the masses $m_Z$, $m_W$, $m_h$, and $m_f$. 
Parameters appearing in the Lagrangian are functions of these input parameters, for example
\be
\ctw = \frac{m_W}{m_Z} \simeq 0.88 \ , \ \ v^2 = \frac{m_W^2}{\pi \, \alpha} \left( 1-\frac{m_W^2}{m_Z^2}  \right) \simeq (243 \, {\rm GeV} )^2 \ ,
\label{vDef}
\ee
where we have used $\alpha^{-1} (m_Z^2)\simeq 129$ \cite{Hoecker:2010qn}. We also have
\be
g=\frac{e}{\stw}  \ , \ \  y_f=\frac{\sqrt{2}\, m_f}{ v}   \ , \ \   \lambda =  \frac{m_h^2}{2 v^2 } \ , \ \   \mu^2 =  \frac{m_h^2}{2 } \ .
\ee

In our calculations, we need the EoM for the Higgs field,
\bea
\partial^\mu \partial_\mu H &=&  
\frac{g^2}{2}  H  \, W^{+\mu} \, W^-_\mu + \frac{g^2}{4\ctw^2}  H  \, Z^{\mu} Z_\mu \nonumber \\
&-& \frac{y_f}{\sqrt{2}}  \bar f f  -\lambda \left( H^2 - v^2 \right) H  \ ,
 \label{eomH}
\eea
and the EoMs for the gauge bosons in the $W^a_\mu, B_\mu$ basis,
\bea
D^\nu  \cW^a_{\mu \nu} &=&  \partial^\nu \cW^a_{\mu \nu}  + g \epsilon^{abc} W^{b \nu} \, \cW^c_{\mu \nu} \nonumber \\
&=& \frac{g}{4}  (g W^a_\mu - g' \delta^{a3} B_\mu) H^2 + g J^a_{ \mu}  \ , \nn \\
\partial^\nu B_{\mu \nu} &=&  - \, \frac{g'}{4}  (g W^3_\mu - g' B_\mu) H^2 + g' J^Y_{ \mu}  \ ,
 \label{eomV}
\eea
where $\cW^a_{\mu \nu}$ and $B_{\mu \nu}$ are the $SU(2)_L$ and $U(1)_Y$ field strengths.

\section*{Appendix B}

The first time a complete basis of $d=6$ operators was presented in the literature was in 
\cite{Grzadkowski:2010es}. The basis  consists of 59 operators (for one family),
with 53 CP-even and 6 CP-odd. We now show that our counting and our basis are consistent with 
the basis in \cite{Grzadkowski:2010es}. We start with the CP-even operators. 

Of the 53 CP-even operators in the basis in \cite{Grzadkowski:2010es}, we do not consider the operator $\cG^3$ 
-the analogous of \eq{3W} with the $SU(3)$ field strength- because it contains exclusively gluon fields.
Also, we do not consider the 25 four-fermion operators, because they do not play a role in our analysis. 
This leaves us with $53-1-25=27$ operators, which is the same number that we have, see \eq{NCPeven}.

We demonstrate now that the two basis of 27 operators are equivalent.
To do it, we follow the classification nomenclature of \cite{Grzadkowski:2010es}, see their Table 2,
and compare with our set of operators. We identify operators that differ only in numerical factors, couplings and signs.

In the group $X^3$, only ${\cal O}_{3W}$ matters. 
Other operators in this group are CP-odd and/or contain only gluons. The operator ${\cal O}_{3W}$ is in our basis, see \eq{3W}.
There are three operators in the group $\phi^6$ and $\phi^4 D^2$: 
${\cal O}_{6}$,  which in our list is \eq{first6dimA}, and two which are not in our basis,
\be
{\cal O}_T=
\frac{1}{2} (\H^\dagger {\lra{D}_\mu} \H)^2 \ \  , \ \ {\cal O}_H = \frac{1}{2}(\partial^\mu |\H|^2)^2  \ .
\label{OT}
\ee

Next group in Table 2 is $\psi^2 \phi^3$. All three operators in this group  are in our basis, they are
${\cal O}_{y_f}$  in \eq{first6dimB}. In the group $X^2 \phi^2$ there are four CP-even operators. 
Three are in our list: ${\cal O}_{GG}, {\cal O}_{BB}, {\cal O}_{WW}$, see \eq{first6dimA}, but not the fourth one:
\be
{\cal O}_{WB}={g}^{\prime}g \,  ( \H^\dagger \sigma^a \H ) \,  \cW^a_{\mu\nu} B^{\mu\nu}  \ .
\label{OWB}
\ee

The dipole group in \cite{Grzadkowski:2010es}, denoted by $\psi^2 X \phi$, contains eight operators. We have these eight dipole operators, 
${\cal O}_{DV}^f$ in \eq{second6dimC}. Finally in the group $\psi^2 \phi^2 D$ in Table 2 of \cite{Grzadkowski:2010es} there are eight operators. 
Seven of them correspond to our seven operators in 
\eq{second6dimB} and the eighth is our \eq{second6dimD}.

As we said, the total number of $CP$-even operators extracted from \cite{Grzadkowski:2010es} which are relevant for our study is 27,
which is precisely the number we have.
We have seen that we have all of them except the three operators in \eq{OT} and \eq{OWB}.
Instead of these three we have  
\be
{\cal O}_{r} \ , \ \ \ {\cal O}_B - {\cal O}_W  \ , \ \ \  {\cal O}_{HB} \ .
\label{Ours}
\ee

The way to relate our
three operators in \eq{Ours} to the three operators in  \eq{OT} and \eq{OWB} is using 
field redefinitions, as explained for example in \cite{Elias-Miro:2013mua}.
We can use the relations
\bea
 {\cal O}_{H} &\leftrightarrow & -  \, {\cal O}_{r}  + \Lambda_1({\cal O}_{y_f}, {\cal O}_6) \nn \\
 g'^2 \, {\cal O}_T  &\leftrightarrow&  -2 \, ({\cal O}_B -  {\cal O}_W) + \Lambda_2({\cal O}_H, {\cal O}_6, {\cal O}_{y_f},
 {\cal O}_L^{(3)\, f}, {\cal O}^f_L, {\cal O}^f_R) \nn \\
 \frac{1}{4} {\cal O}_{WB} &\leftrightarrow & -  \, {\cal O}_{HB}  +  ({\cal O}_B -  {\cal O}_W) + 
\Lambda_3({\cal O}_{BB}, {\cal O}_H, {\cal O}_6, {\cal O}_{y_f},
 {\cal O}_L^{(3)\, f}, {\cal O}^f_L, {\cal O}^f_R)
\eea
The symbol $\leftrightarrow$ means here that the rhs and lhs can
be traded one by the other.  We have introduced $\Lambda_{1,2,3}$ which are well-defined linear functions of operators 
that are common to both basis. For the explicit form of these functions see
\cite{Elias-Miro:2013mua}.

With that we have shown that the two basis are equivalent in the CP-even sector.
Let us finally examine the CP-odd operators. In \cite{Grzadkowski:2010es} there 5 such operators, if we exclude the CP-odd triple gluon
operators, as we have done. Four of such operators are common: ${\cal O}_{B \widetilde  B}$, ${\cal O}_{W \widetilde  W}$,
 ${\cal O}_{G \widetilde  G}$, ${\cal O}_{3\widetilde W}$, see \eq{Oodd1} and \eq{Oodd2}, but one is different. While they have the operator
\be
{\cal O}_{W\widetilde B}={g}^{\prime}g \,  ( \H^\dagger \sigma^a \H ) \,  \cW^a_{\mu\nu} \widetilde B^{\mu\nu}  \ .
\label{OWtildeB}
\ee
we have the operator ${\cal O}_{H\widetilde B}$ in \eq{Oodd2}. Both operators can be related by
\be
{\cal O}_{W\widetilde B} \leftrightarrow  -  \, 4 \, {\cal O}_{H \widetilde  B} - {\cal O}_{W \widetilde  W} 
\ee

\end{document}